\documentclass[fleqn,usenatbib]{mnras}
\usepackage{longtable}
\usepackage{multirow}

\usepackage[T1]{fontenc}

\DeclareRobustCommand{\VAN}[3]{#2}
\let\VANthebibliography\thebibliography
\def\thebibliography{\DeclareRobustCommand{\VAN}[3]{##3}\VANthebibliography}

\newcommand{\sigmaHI}{$\sigma_{_{\rm HI}}$}
\newcommand{\sigmaHtwo}{$\sigma_{_{\rm H_2}}$}

\newcommand{\BB}{$^{\scriptstyle 3\mathrm{D}}$B{\sc arolo}}
\newcommand{\hi}{\ifmmode{\rm HI}\else{H\/{\sc i}}\fi}

\usepackage{graphicx}	
\usepackage{amsmath}	
\usepackage{amssymb}	

\title[Gas scale heights, mass models, and scaling relations of nearby disc galaxies]{The impact of gas disc flaring on rotation curve decomposition and revisiting baryonic and dark-matter relations for nearby galaxies}

\author[Pavel E. Mancera Pi\~na et al.]{Pavel E. Mancera Pi\~na$^{1,2}$\thanks{e-mail: pavel@astro.rug.nl}, Filippo Fraternali$^{1}$, Tom Oosterloo$^{2,1}$, Elizabeth A. K. Adams$^{2,1}$,\newauthor Enrico di Teodoro$^{3,4}$, Cecilia Bacchini$^{5}$, and Giuliano Iorio${^{5,6}}$\\
\\
$^{1}$ Kapteyn Astronomical Institute, University of Groningen, Landleven 12, 9747 AD, Groningen, The Netherlands\\
$^{2}$ ASTRON, Netherlands Institute for Radio Astronomy, Postbus 2, 7900 AA Dwingeloo, The Netherlands\\
$^{3}$ Department of Physics \& Astronomy, Johns Hopkins University, Baltimore, MD 21218, USA \\
$^{4}$ Space Telescope Science Institute, 3700 San Martin Drive, Baltimore, MD 21218, USA \\
$^{5}$ INAF-Osservatorio Astronomica di Padova, Vicolo dell'Osservatorio 5, I35122 Padova, Italy\\
$^{6}$Physics and Astronomy Department Galileo Galilei, University of Padova, Vicolo dell'Osservatorio 3, I-35122, Padova, Italy
}

\pubyear{2022}
\usepackage{newtxtext,newtxmath}

\begin{document}
\label{firstpage}
\pagerange{\pageref{firstpage}--\pageref{lastpage}}
\maketitle

\begin{abstract}

Gas discs of late-type galaxies are flared, with scale heights increasing with the distance from the galaxy centres and often reaching kpc scales. We study the effects of gas disc flaring on the recovered dark matter halo parameters from rotation curve decomposition. For this, we carefully select a sample of 32 dwarf and spiral galaxies with high-quality neutral gas, molecular gas, and stellar mass profiles, robust H\,{\sc i} rotation curves obtained via 3D kinematic modelling, and reliable bulge-disc decomposition. By assuming vertical hydrostatic equilibrium, we derive the scale heights of the atomic and molecular gas discs and fit dark matter haloes to the rotation curves self-consistently. 
We find that the effect of the gas flaring in the rotation curve decomposition can play an important role only for the smallest, gas-dominated dwarfs, while for most of the galaxies the effect is minor and can be ignored.
We revisit the stellar-- and baryon--to--halo mass relations ($M_\ast-M_{200}$ and $M_{\rm bar}-M_{200}$). Both relations increase smoothly up to $M_{200} \approx 10^{12}~\rm{ M_\odot}$, with galaxies at this end having high $M_\ast/M_{200}$ and $M_{\rm bar}/M_{200}$ ratios approaching the cosmological baryon fraction. At higher $M_{200}$ the relations show a larger scatter. Most haloes of our galaxy sample closely follow the concentration--mass ($c_{200}-M_{\rm 200}$) relation resulting from N-body cosmological simulations. Interestingly, the galaxies deviating above and below the relation have the highest and lowest stellar and baryon factions, respectively, which suggests that the departures from the $c_{200}-M_{\rm 200}$ law are regulated by adiabatic contraction and an increasing importance of feedback.
\end{abstract}

\begin{keywords}
galaxies: evolution -- galaxies: fundamental parameters -- galaxies: kinematics and dynamics -- galaxies: ISM -- galaxies: structure -- cosmology: dark matter
\end{keywords}



\section{Introduction}
About 50 years have passed since the first set of works aiming to use rotation curve decomposition to study, and provide evidence for, the dark matter content of galaxies (e.g. \citealt{freeman1970,roberts1973,shostak1973,bosma1978,rubin1980,vanalbada86,begeman}, see also \citealt{bertone_DMreview} for a historical review). The main idea is well known: provided the distribution of stars and gas within a galaxy, one can compare the contribution from these baryonic components to the gravitational potential against the total potential traced by the observed rotation curve of the galaxy, and infer the mass distribution of the host dark matter halo. Significant work has also been done using rotation curves decomposition to test alternative theories to dark matter (e.g. \citealt{mond,kent1987,mcgaugh_mondLSB,mond_famaey_review}). 


The derivation of mass models has become significantly more accurate and detailed than in the first studies half a century ago. Nowadays, mass models use high quality data with high resolution kinematic measurements (typically H\,{\sc i} data) and accurate near-IR photometry to study the dynamics of relatively large galaxy samples (e.g. \citealt{deblok08,frank2016,sparc}), using robust statistical techniques such as Bayesian inference (e.g. \citealt{readAD,postinomissing}), and testing a variety of dark matter haloes and even different types of dark matter (e.g. \citealt{ren_SIDM_SPARC,li_massmodels,zentner2022}). These models are used on a daily basis for a variety of applications on galaxy evolution and near-field cosmology.

A limitation of almost all the mass models available in the literature, is that they assume that the gas (and sometimes also the stars) is distributed in razor-thin discs. Yet, it has been well established observationally that gas discs are not only thick, but they are also flared, with a scale height increasing with galactocentric distance, as expected for systems in vertical hydrostatic equilibrium (e.g. \citealt{romeo1992,olling1995,yim2014,bookFilippo}). 

Other than in the Milky Way (e.g. \citealt{nakanishi2003,karlberla2008, marasco2011,marasco2017}), the flaring of gas discs can be measured directly only in highly-inclined galaxies, although this is challenging as there are degeneracies between the thickness of the discs and the inclination of galaxies, warps along the line of sight, or the presence of layers of extra-planar gas (e.g. \citealt{kerr1957,renzo1979,merrifield1992,oosterloo_2007,obrien2010_results,yim2014}). Notwithstanding, the scale height of any galaxy can also be obtained by solving the equations of vertical hydrostatic equilibrium (i.e. assuming that the gas pressure is in balance with
the gravitational pull) given the galactic potential (see for instance \citealt{vanderkruitNGC891,romeo1992,olling1995,iorio_phd, ceciVSFL,ceci_dwarfs,patra_HIdwarfs,patra_HIspirals}). In general, evidence shows clearly that in both dwarfs and massive late-type galaxies, the scale heights of gas discs at the outermost observed radii increase typically by a factor $2-10$ with respect to the innermost regions, often reaching values $\gtrsim 1$~kpc. 

Understanding the real geometry of gas discs has important implications for our understanding of star formation, turbulence, and feedback processes (e.g. \citealt{romeo1992,ceciVSFL,utomo2019,ceci_turbulence}). The flaring of the discs can also play a role when deriving the rotation curve decomposition and mass models of galaxies. Given that the circular speed profiles of discs will depend on their gravitational potential and more specifically on the radial acceleration at the midplane, a varying scale height will produce a different circular speed than razor-thin discs and even than thick discs with constant thickness (e.g. \citealt{binney}). The effects are systematic and should be quantified.

In this paper, we aim to investigate how significantly affected mass models are when taking into account the flared geometry of the gas discs. To the best of our knowledge, the effect of gas flaring in the mass models has not been studied in the literature in a systematic way for a sample of galaxies (see e.g. \citealt{olling1996a} for an individual case). So far, mass models that assume razor-thin discs have been used to derive the scale height of the H\,{\sc i} and H$_2$ discs, but the impact of this on the recovered dark matter halo parameters themselves is yet to be explored. It is important to understand and quantify this, specially with the advent of large and systematic studies providing mass models. In this work, we start filling this gap by self-consistently deriving the gas flaring and dark matter halo parameters for a sample of galaxies.\\ 

\noindent
Accurate computations of scale heights require detailed kinematic modelling of interferometric H\,{\sc i} and CO data, together with robust bulge-disc decomposition. As we detailed below, we have collected such data. This also allow us to revisit important scaling relations, namely the stellar- and baryon--to--halo mass relation, and the dark matter halo concentration--mass relation. While these relations have been recently explored using mass models of large galaxy samples (e.g. \citealt{postinomissing,li_massmodels}), some questions remain open. For instance, $\textit{i)}$ the shape of the stellar--to--halo mass relation at the lowest and highest stellar mass regime is not fully established (e.g. \citealt{moster2013,posti_galaxyhalo}), \textit{ii)} the baryon--to--halo mass relation remains somewhat unexplored as mass models do only rarely include both the atomic and molecular gas content, and \textit{iii)} it has been claimed that the dark matter concentration--mass relation from N-body cosmological simulations (e.g. \citealt{duttonmaccio2014,ludlow2014}) is not followed by the data unless imposed as a prior (e.g. \citealt{katz2017,li_massmodels}). In addition to this, it is unknown whether the flaring of the discs can affect these relations. Gaining insight into these scaling laws is essential in the quest of understanding galaxy evolution. Besides encoding dynamical information, these relations are subject to baryonic physics and thus we can use them to constrain processes like feedback and adiabatic contraction, which are expected to impact the properties of the dark matter haloes (e.g. \citealt{mo_book,bookFilippo}). We will therefore take advantage of our analysis and data to revisit these relations.\\

\noindent
The structure of this work is as follows. In Sec.~\ref{sec:data} we present our galaxy sample, explain its selection criteria, and describe the type of data used for each of our galaxies. In Sec.~\ref{sec:methods} we detail our methodology to derive the gas discs scale heights and mass models for each galaxy. In Sec.~\ref{sec:results} we present our main findings, showing our mass models and scale heights. In Sec.~\ref{sec:discussion} we discuss the implications of our results for the scaling relations between halo mass and concentration, stellar mass, and baryonic mass; we also discuss the main caveats of our study and compare with previous works. Finally, in Sec.~\ref{sec:conclusions} we summarise our main findings.



\section{Sample selection and data}
\label{sec:data}

To investigate the effects of the gas thickness on mass models, circular speeds, gas velocity dispersions, and mass profiles are needed. This makes it essential to select galaxies with high-quality data, mapping their gas and stellar distributions. In the case of the gas surface density profiles, deep, extended, and high-spatial resolution data (typically H\,{\sc i} or CO emission-line observations) are needed. For the stellar component, near-IR data is desirable to trace the underlying stellar discs. Concerning the kinematics, the H\,{\sc i} and CO data also allow to obtain the gas rotation velocity and velocity dispersion, which can then be used to obtain the circular speed. As we describe below, our sample consists of galaxies with these kind of data and with detailed kinematic models available in the literature. 

\begin{figure*}
    \centering
    \includegraphics[scale=0.48]{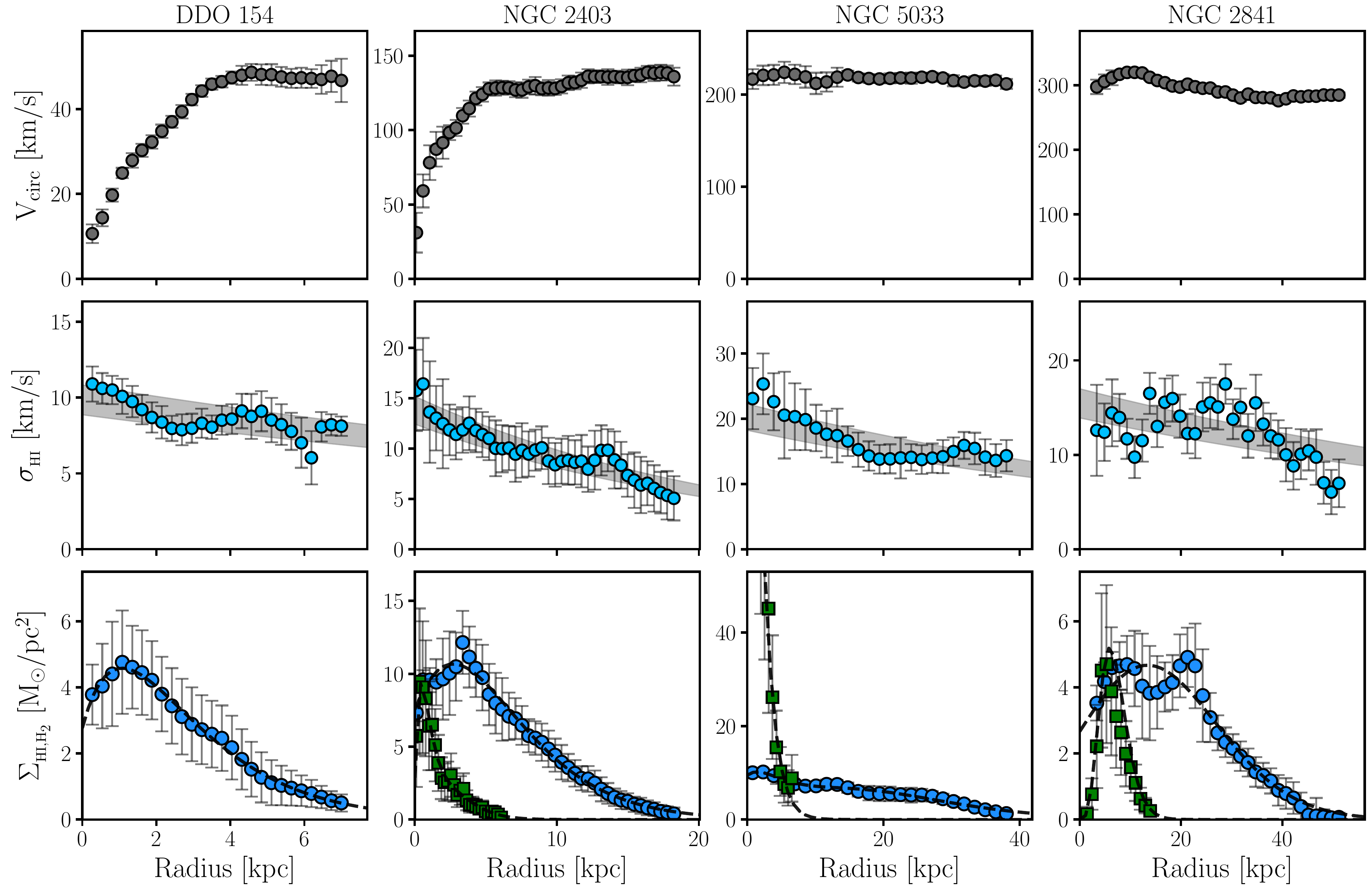}
    \caption{Data for four representative galaxies in our sample spanning our full circular speed range. \textit{Top:} Circular speed profiles. \textit{Middle:} H\,{\sc i} velocity dispersion profiles. The gray band shows an exponential fit to the data (blue) with an uncertainty of $\pm$ 10 per cent in its normalisation. \textit{Bottom:} H\,{\sc i} (blue circles) and H$_2$ (green squares) surface density profiles, including helium correction. The dwarf DDO~133 has no available CO data and it likely hosts no significant H$_2$ reservoirs. The curves on top of the data show the functional forms fit to obtain the corresponding contributions to the total circular speed (see Sec.~\ref{sec:circular_speeds}). In all the panels for NGC~2403 and NGC~2841 we plot one data point for every two, except for the H$_2$ surface density profiles.}
    \label{fig:show_data}
\end{figure*}

\subsection{Dwarf galaxies}
\subsubsection{Kinematics}
We selected a subsample of the LITTLE THINGS galaxies \citep{hunterLT} studied by \citet{iorio}. Those authors provide the H\,{\sc i} velocity dispersion (\sigmaHI{}) as well as the circular speed ($V_{\rm c}$) of the galaxies after correcting the observed rotation velocities ($V_{\rm rot}$) for asymmetric drift (see Sec. 4.3 in \citealt{iorio}). These parameters\footnote{All the kinematic data on LITTLE THINGS galaxies studied in \citet{iorio} is publicly available at \url{https://www.filippofraternali.com/downloads}} were derived using \BB{} \citep{barolo}\footnote{\url{https://editeodoro.github.io/Bbarolo/}}, a software that fits non-parametric tilted ring models to emission-line data cubes. Effectively, \BB{} creates 3D realizations of rotating gas discs and chooses the model that better reproduces the emission-line data, velocity channel by velocity channel, after taking into account the beam (PSF) of the observations through a convolution operation. For large disc galaxies traditional 2D methods fitting the velocity field (e.g. \citealt{deblok08,frank2016}) and \BB{} yield similar results. However, dwarf galaxies (and low-resolution observations in general) may suffer from beam smearing, which can significantly modify the shape of the velocity fields (e.g. \citealt{begeman,swatersPhD,barolo}), making 2D methods unreliable to recover the intrinsic kinematics. The 3D approach of \BB{} is thus mostly necessary when deriving the kinematics of dwarfs, as it largely mitigates the effects of beam smearing. Moreover, 3D methods have also the advantage of constraining simultaneously the rotation of the galaxy as well as the gas velocity dispersion.

Out of the 17 systems studied by \citet{iorio}, we remove the galaxies DDO~47, DDO~50, DDO~53, DDO~101, DDO~133, DDO~216, and NGC~1569. As discussed in detail by \cite{read2017}, the above galaxies are subject of concern regarding their dynamical state, distance estimation, or inclination determination, so we prefer to keep them out of our analysis, ending up with ten dwarfs. It is worth mentioning that all the galaxies have inclination angles larger than 30 degrees, which is important as uncertainties in low inclination angles translate into big uncertainties in the deprojected rotation velocities. 

\subsubsection{Gas and NIR optical surface densities}

The H\,{\sc i} surface densities are taken directly from \cite{iorio}, and we only apply a multiplicative factor of 1.36 to account for the presence of helium. Stellar 3.6~$\mu$m surface brightness profiles are available from the S4G Survey (\citealt{galexs4g}, see also \citealt{sheth2010}) and from \citet{zhang}. Molecular gas is not detected in these dwarfs (e.g. \citealt{leroy,bigiel}) as it often happens with low-mass galaxies, probably due to their low metallicities (see \citealt{hunt2015} and references therein). Recently, \citet{hunter_H2} attempted to infer the molecular gas content of LITTLE THINGS galaxies based on their FUV emission. Those authors report an average H$_2$--to--H\,{\sc i} fraction of about 0.2. Given that these are not direct measurements of H$_2$ and that the H$_2$--to--H\,{\sc i} fraction of \citet{hunter_H2} is relatively small, in our analysis we do not attempt to include the molecular gas contribution in dwarf galaxies.

\subsection{Spiral galaxies}
\subsubsection{Kinematics}
To define our spiral galaxy sample, we built on the work by \citet{enrico_radialmotions}, who also used \BB{} to derive the H\,{\sc i} kinematics of a sample of nearby spiral galaxies. Broadly speaking, and as detailed below, we chose their best galaxies in terms of spatial resolution, undisturbed kinematics, available bulge-disc NIR photometric decomposition, and ancillary CO observations. 

Among all the galaxies from \citet{enrico_radialmotions}, we make a first selection cut and keep only those with a spatial resolution better than 1~kpc, to ensure good sampling of the rotation curve and velocity dispersion profiles. Next, we selected against galaxies with inclinations below $30^\circ$. After this, we also eliminate the galaxies NGC~3031, NGC~3521, M83,  which met the above criteria but show very perturbed kinematics or interactions with neighbors and thus it is not granted that their gas kinematics fully trace their potential wells.

Regarding the kinematics of the galaxies, we took the values from \cite{enrico_radialmotions} for $V_{\rm rot}$\footnote{For a few galaxies we have removed a couple of data points near their centres where it is unclear whether there is actually emission (e.g. the centre of NGC~2841), or on the outer regions if the galaxies show a strong warp or an asymmetric rotation curve (e.g. NGC~3621).} and converted them to $V_{\rm c}$ following the same asymmetric drift correction prescriptions detailed in \cite{iorio}. This correction is extremely small and rather negligible for this sample of massive galaxies, but we undertook it for consistency with the LITTLE THINGS dwarfs (where the correction is significant in some cases). An additional kinematic parameter that we will need for the mass models and the derivation of the scale height of the H$_2$ discs is their velocity dispersion (see Section~\ref{sec:thickness}). The rotation curves of H\,{\sc i} and H$_2$ (as traced by the CO line) show a very good agreement \citep{frank2016,ceci_turbulence} between each other. In contrast, the H$_2$ velocity dispersion (\sigmaHtwo{}) and \sigmaHI{} are not the same. \citet[][see also e.g. \citealt{mogotsi2016}]{ceci_turbulence} found that the ratio between \sigmaHtwo{} and \sigmaHI{} among late-type galaxies has a median value of 0.6, i.e. \sigmaHtwo{} $\approx$~0.6 \sigmaHI{}. In principle, one could derive the velocity dispersion of the molecular gas from the CO data (see below). However, this would require the CO kinematic modelling and for some galaxies the CO emission is patchy and the measurement of \sigmaHtwo{} is not well constrained and has large uncertainties. For the sake of homogeneity and to avoid these complications, we use the empirical results of \citet{ceci_turbulence} and we assume \sigmaHtwo{}$= 0.6$ \sigmaHI{}; we have checked that reasonable variations of this equality do not impact our results on the mass models significantly. 

\subsubsection{Gas surface densities}

The H\,{\sc i} surface density profiles are taken from \cite{enrico_radialmotions}, and we apply the helium multiplicative factor of 1.36. In the case of the molecular gas, we derive H$_2$ profiles from a set of CO$(J=2-1)$ and CO$(J=1-0)$ total intensity maps; the existence of CO ancillary observations was one of the main criteria when selecting our sample. For this, we take advantage of following archival observations. The CO total intensity maps of the galaxies NGC~253, NGC~3351, NGC~3621, NGC~4535, and NGC~4536, come from the PHANGS-ALMA survey \citep{phangs_data,phangs_reduction}. The maps of NGC~2403, NGC~2841, NGC~3198, NGC~4559, NGC~4725, NGC~4736, and NGC~5055 come from the HERACLES survey \citep{heracles}. The maps of NGC~5005 and NGC~5033 come from the BIMA project \citep{bima}, and of NGC~4651 and NGC~4698 from the VERTICO survey \citep{vertico}. The CO map of NGC~3486 comes from \cite{carma}. As the CO map of NGC~3675 is not available, we extracted the published CO profile shown in \cite{fcrao} using a digitizing software \citep{digitizer}. Finally, the galaxies NGC~1313 and NGC~3992 have no significant CO detection in the available data (\citealt{bajaja1995} and \citealt{bima}, respectively) and CO data for NGC~3898 and NGC~5350 do not seem to exist. These last two galaxies have, however, high circular speeds (about 250~km/s and 185~km/s, see Table~\ref{tab:sample}) dominated in the inner regions by massive bulges and stellar discs, so we do not expect the lack of CO data to affect their rotation curve decomposition. The collected data have different sensitivity and spatial resolution, but overall they allow us to trace reasonably well the CO emission in the central regions of our sample.

Using the CO maps we build azimuthally averaged intensity maps using the same average geometric parameters as for the H\,{\sc i} discs\footnote{We note that there is an excellent agreement between the geometric parameters of the H\,{\sc i} discs \citep{enrico_radialmotions} and the reported values for the CO discs (e.g. \citealt{phangs_data,vertico}).}, and we convert the CO intensities $I_{\rm CO}^{(2-1)}$ to H$_2$ mass surface densities using
\begin{equation}
\label{eq:H2mass}
\dfrac{\Sigma_{\rm {H_2}}}{M_\odot\ \rm{pc}^{-2}} = \dfrac{\alpha_{_{\rm{CO}}}^{(1-0)}}{M_\odot\ \rm{pc}^{-2}}
      \dfrac{1}{R_{21}} \dfrac{I_{\rm CO}^{(2-1)}}{\rm{K\ km/s}},
\end{equation}
where $\alpha_{_{\rm{CO}}}^{(1-0)}$ is the CO--H$_2$ conversion factor (including the helium correction factor of 1.36) and $R_{21}$ the CO($2-1$)--to--CO($1-0$) intensity line ratio, for which we assume a value of 0.7 (see e.g. \citealt{leroy2013,sandstrom2013,phangs_data})\footnote{The CO data for NGC~3486, NGC~3675, NGC~5005, and NGC~5033 are for the transition $J=1-0$, and so we take $R_{21}$=1 in Eq.~\ref{eq:H2mass}.}.
While a handful of our galaxies have available $\alpha_{_{\rm{CO}}}^{(1-0)}$ measurements from \cite{sandstrom2013}, the majority of our sample does not have available $\alpha_{_{\rm{CO}}}^{(1-0)}$ values. This is not surprising given the difficulty in estimating $\alpha_{_{\rm{CO}}}^{(1-0)}$ precisely.
For consistency, for all our spiral galaxies we adopt $\alpha_{_{\rm{CO}}}^{(1-0)} = 4.35$, the Milky Way reference value (e.g. \citealt{bolatto_review} and references therein), which is also in reasonable agreement with the average value measured by \citet{sandstrom2013} in a sample of nearby spiral galaxies. \cite{sandstrom2013} also found radial variations on the $\alpha_{_{\rm{CO}}}^{(1-0)}$ of some galaxies in their sample. However, such variations are not always present nor seem to correlate significantly with other galaxy properties, and while most galaxies have somewhat lower values in the central regions, the scatter is large and trends with galactocentric distance are unclear. Given that molecular gas is not the dominant mass component of our galaxies at any radii we do not expect the potential uncertainties in $\alpha_{_{\rm{CO}}}^{(1-0)}$ to have a strong effect in our rotation curve decomposition. In fact, this has already been shown by \citet{frank2016}. 



\subsubsection{NIR surface brightness and bulge-disc decomposition}

Given that the bulge component will be relevant when deriving our mass models, we refined our selection by requiring that our spiral galaxies have available bulge-disc decompositions from the S4G data coming from \cite{salo2015}. Those authors performed a detailed decomposition of the 3.6~$\mu$m surface brightness profiles of the galaxies fitting an exponential disc and a S\'ersic profile, and we stick to their decomposition for the sake of consistency. We also select against galaxies with a significant bar or central PSF-like component as determining their kinematics and potentials can be ambiguous (e.g. NGC~4548). For three galaxies in our sample (NGC~2403, NGC~3198, NGC~3621) we do not use the S4G bulge-disc decomposition. The first and the last are not in the sample of \cite{salo2015}, while the fit for NGC~3198 is not satisfactory (e.g. the S\'ersic index is 10, the maximum allowed value by the fitting routine of \citealt{salo2015}). These three galaxies have, however, 3.6~$\mu$m S4G and optical surface brightness profiles without evidence of a bulge \citep{kent1987,munozmateos}. Because of this, and since the galaxies have exquisite H\,{\sc i} and CO data and are prototypical well-studied nearby galaxies, we include them in our final sample.

\subsection{Final galaxy sample}

In the end, our final sample consists of ten dwarf galaxies with $15 \lesssim V_{\rm c}/(\rm{km/s}) \lesssim 60$ and 22 more massive spirals with $V_{\rm c}$ as high as $\sim 300$~km/s. Table~\ref{tab:sample} lists our final galaxy sample, and gives the distance to the galaxies (taken form \citealt{enrico_radialmotions} and \citealt{paperIBFR}), their morphological type taken from NED\footnote{The NASA/IPAC Extragalactic Database (NED) is operated by the Jet Propulsion Laboratory, California Institute of Technology, under contract with the National Aeronautics and Space Administration.}, and their characteristic circular speed at the outer radii $V_{\rm c,out}$. For the spiral galaxies this characteristic speed is computed following the algorithm of \citet[][see their Eq. 1 and 2]{lelliBTFR}, i.e., by requiring that the circular speeds are flat within $\sim 5$ per cent over at least the last three measured points (but often many more points are included). The reported uncertainties correspond to the standard deviation in the values considered to estimate the flat speed. In the case of the dwarfs, where some of them have fewer than three points consistent with flat rotation, we use the mean outer velocity reported by \cite{iorio}. Fig.~\ref{fig:show_data} shows the circular speeds, H\,{\sc i} velocity dispersions, and gas surface densities for four representative galaxies in our sample spanning our full circular speed range.

\begin{table}
\caption{Galaxy sample used in this work. The first column gives the names of the galaxies, the second column their distance, the third column their morphological classification, and the fourth column their characteristic circular speed at the outermost radii.}
\label{tab:sample}
\begin{center} 
\begin{tabular}{lccc}
	\hline \noalign{\smallskip}
 Name & Distance   & Morphology & $V_{\rm c,out}$  \\ 
      & [Mpc] &    & [km/s] \\ \noalign{\smallskip}
   \hline \noalign{\smallskip}
   \multicolumn{4}{c}{Dwarf galaxies} \\ \noalign{\smallskip} \hline
CV n I dwA   &  3.6 $\pm$ 0.2  & Im & 22 $\pm$ 4 \\ 
   DDO 52    & 10.3 $\pm$ 0.5  & Im & 51 $\pm$ 6 \\ 
   DDO 87    &  7.4 $\pm$ 2.2  & Im & 50 $\pm$ 9 \\ 
   DDO 126   &  4.9 $\pm$ 0.4  & IBm & 39 $\pm$ 3 \\ 
   DDO 154   &  3.7 $\pm$ 0.1  & IB(s)m & 47 $\pm$ 5 \\ 
   DDO 168   &  4.3 $\pm$ 0.3  & IBm & 56 $\pm$ 7 \\ 
   DDO 210   &  0.9 $\pm$ 0.1  & IB(s)m & 16 $\pm$ 10 \\ 
   NGC 2366  &  3.4 $\pm$ 0.3  & SB(s)b & 58 $\pm$ 5 \\ 
   UGC 8508  &  2.6 $\pm$ 0.2  & IAm & 34 $\pm$ 6 \\ 
   WLM       &  1.0 $\pm$ 0.1  & IB(s)m  & 39 $\pm$ 3 \\ \noalign{\smallskip} 
   \hline
\multicolumn{4}{c}{Spiral galaxies} \\ \noalign{\smallskip} \hline
 NGC 0253   &   3.6 $\pm$ 0.3  & SAB(s)c & 198 $\pm$ 2 \\ 
 NGC 1313   &   4.2 $\pm$ 0.4  & SB(s)d  & 127 $\pm$ 3 \\ 
 NGC 2403   &   3.2 $\pm$ 0.3  & SAB(s)cd & 136 $\pm$ 2 \\ 
 NGC 2841   &  14.1 $\pm$ 1.9  & SA(r)b & 286 $\pm$ 6 \\ 
 NGC 3198   &  13.8 $\pm$ 1.1  & SB(rs)c & 150 $\pm$ 3 \\ 
 NGC 3351   &  10.5 $\pm$ 0.8  & SB(r)b & 177 $\pm$ 4 \\ 
 NGC 3486   &  12.7 $\pm$ 2.8  & SAB(r)c & 155 $\pm$ 3 \\ 
 NGC 3621   &   6.7 $\pm$ 0.5  & SA(s)d & 145 $\pm$ 2 \\ 
 NGC 3675   &  14.4 $\pm$ 3.1  & SA(s)b & 213 $\pm$ 4 \\ 
 NGC 3898   &  22.1 $\pm$ 6.1  & SA(s)ab & 246 $\pm$ 5 \\ 
 NGC 3992   &  17.1 $\pm$ 4.7  & SB(rs)bc & 231 $\pm$ 3 \\ 
 NGC 4535   &  15.8 $\pm$ 1.3  & SAB(s)c & 225 $\pm$ 6 \\ 
 NGC 4536   &  15.2 $\pm$ 1.9  & SAB(rs)bc & 165 $\pm$ 4 \\ 
 NGC 4559   &   8.9 $\pm$ 0.8  & SAB(rs)cd & 123 $\pm$ 3 \\ 
 NGC 4651   &  16.8 $\pm$ 4.6  & SA(rs)c & 185 $\pm$ 3 \\ 
 NGC 4698   &  19.6 $\pm$ 4.2  & SA(s)ab & 214 $\pm$ 5 \\ 
 NGC 4725   &  12.4 $\pm$ 1.1  & SAB(r)ab & 211 $\pm$ 3 \\ 
 NGC 4736   &   4.4 $\pm$ 0.4  & (R)SA(r)ab & 148 $\pm$ 2 \\ 
 NGC 5005   &  18.4 $\pm$ 2.1  & SAB(rs)bc & 269 $\pm$ 6 \\ 
 NGC 5033   &  19.0 $\pm$ 2.7  & SA(s)c & 217 $\pm$ 3 \\ 
 NGC 5055   &   8.9 $\pm$ 0.7  & SA(rs)bc & 184 $\pm$ 6 \\ 
 NGC 5350   &  30.9  $\pm$ 5.5 & SB(r)b & 184 $\pm$ 2 \\  
 \hline
    \end{tabular}
  \end{center}
\end{table}

We would like to highlight that although the galaxies in our sample have data coming from different studies, they all have the same type of data and have been analyzed with the same methods and thus the sample is fairly homogeneous: 1) the H\,{\sc i} kinematics for all the galaxies are derived using \BB{} on data cubes, 2) all the H\,{\sc i} and H$_2$ surface density profiles are based on azimuthal averages of H\,{\sc i} and CO maps, and 3) all the galaxies have 3.6~$\mu$m photometry and are either bulgeless or have available bulge-disc decomposition derived with the same methodology. This sample is not complete, but as similar samples often used in the literature (e.g. \citealt{sparc,anastasia1,paperIBFR}), it is representative of the regularly rotating nearby galaxy population.

\section{Obtaining the disc thickness and dark matter halo}
\label{sec:methods}
The general idea of mass modelling from rotation curve decomposition is that the observed circular speed profile $V_{\rm c}$ is the sum in quadrature of the circular speed profiles of the baryonic and dark matter components
\begin{equation}
\label{eq:massmodel}
    V_{\rm c}^2 = \Upsilon_{\rm d} V_{\rm d}|V_{\rm d}| + \Upsilon_{\rm b} V_{\rm b}|V_{\rm b}| + V_{\rm HI}|V_{\rm HI}| + V_{\rm H_2}|V_{\rm {H_2}}| + V_{\rm DM}|V_{\rm DM}|~,
\end{equation}
with $V_{\rm d}$, $V_{\rm b}$, $V_{\rm HI}$, $V_{\rm H_2}$, and $V_{\rm DM}$ the contributions to the circular speed by the stellar disc, stellar bulge, H\,{\sc i} disc, H$_2$ disc, and dark-matter halo, respectively, and with $\Upsilon_{\rm d}$ ($\Upsilon_{\rm b}$) the disc (bulge) mass-to-light ratio.

Our approach will be to fit $V_{\rm c}$ and obtain the dark matter halo by constraining $V_{\rm DM}$ (assuming a functional form with free parameters, see below) together with $\Upsilon_{\rm d}$ and $\Upsilon_{\rm b}$. The key novelty with respect to previous studies is that the thickness of the gas discs is derived self-consistently using an iterative procedure (and thus changing $V_{\rm HI}$ and $V_{\rm H_2}$). In the rest of this section we explain the general steps followed to obtain our final mass models.

\subsection{From gas and stellar surface densities to circular speed profiles}
\label{sec:circular_speeds}

Once the geometry and surface density of a matter component are known, its circular speed profile can be obtained from its gravitational potential. To derive the circular speed profiles from the surface densities, we use the software \textsc{galpynamics}\footnote{\url{https://gitlab.com/iogiul/galpynamics}} \citep{iorio_phd}. In the case of galactic discs, \textsc{galpynamics} computes the gravitational potential of a given mass distribution of density $\rho$ via numerical integration of the expression (cf. \citealt{cuddeford1993})
\begin{equation}
\label{eq:full_potential}
    \Phi(R,z) = -\dfrac{2\ G}{\sqrt{R}} \int^{\infty}_{-\infty} dl \int^{\infty}_{0} du\ \sqrt{u\ y}\ K(\sqrt{y\ }) \rho(u,l)~, 
\end{equation}
where $R$ and $z$ are the radial and vertical coordinates, $G$ is the gravitational constant, $K$ the complete elliptical integral of the first kind, and $y$ defined as:
\begin{equation}
    y = \dfrac{2}{1+x},\ \textrm{with}\ x = \dfrac{R^2 + u^2 +(z-l)^2}{2Ru} ~.
\end{equation}
The circular speed is then obtained from the derivative of the potential evaluated at the midplane
\begin{equation}
    V_{\rm c}^2(R) = R \left[ \dfrac{\partial \Phi(R,z)}{\partial R} \right]_{z=0}~.
    \label{eq:vc_potential}
\end{equation}
In the case of a spherical distribution the value of $V_{\rm c}$ is simply $\sqrt{G M(r)/r}$. While deriving $V_{\rm c}$ for a given component, \textsc{galpynamics} uses functional forms to describe its surface density profile. Because of this, we fit the observed gas and stellar surface densities with functional forms as described below.

\subsubsection{Dwarfs: HI distribution}

The surface density profiles of the H\,{\sc i} discs of the dwarfs are well described by a function of the form
\begin{equation}
\label{eq:Frat_disc}
    \Sigma_{\rm HI}(R) = \Sigma_{0,\rm HI}~e^{-R/{\rm R_{1}}} (1+R/{\rm R_{2}})^\alpha~,
\end{equation}
with $\Sigma_0$, $R_1$, $R_2$ and $\alpha$ the fitting parameters. This function has a characteristic shape that mimics very well the surface density profiles observed in gas-rich galaxies, which often show a plateau or central depression in the inner regions, then they increase, and then fall almost exponentially in the outskirts (e.g. \citealt{swatersPhD,martinssonHI,wang16}), as illustrated in Fig.~\ref{fig:show_data} for DDO~133, NGC~2403, and NGC~2841.

Regarding the vertical distribution, while in a first step we assume the disc is razor-thin, the final scale height and its flaring are determined at the same time as the dark matter halo parameters, as we describe in Sec.~\ref{sec:themethod}.

\subsubsection{Dwarfs: stellar discs}

All our dwarf galaxies have stellar disc distributions that are very well fitted by exponential discs, with minimal departures. Therefore, we fit the stellar 3.6~$\mu$m profiles as exponential discs following
\begin{equation}
\label{eq:exp_disc}
    \Sigma_{\ast}(R) = \Sigma_{\rm 0,\ast}~e^{-R/R_{\rm d}}~,
\end{equation}
with $\Sigma_{\rm 0,\ast}$ and $R_{\rm d}$ the stellar central surface density and disc scale length, respectively.

To fully define the stellar disc distribution, we also need to specify the vertical structure. We assume that the stellar disc follows a sech$^2$ profile along the vertical direction, and that it has a constant thickness $z_{\rm d} = 0.196 R_{\rm d}^{0.633}$, as found in low-inclination star forming galaxies (e.g., \citealt{bershady_thickness}; see also \citealt{vanderkruit2011}). With this, the shape of $V_{\rm d}$ for the dwarfs is fully determined, and during our fitting technique we will obtain $\Upsilon_{\rm d}$ (i.e. its normalization). We remind the reader that there are no bulges present in our dwarfs, neither do the galaxies have significant H$_2$, and thus in practice $V_{\rm b} = V_{\rm H_2} = 0$.

\subsubsection{Spirals: H\,{\sc i} and H$_2$ distributions}
As for the dwarf galaxies, we fit some of the H\,{\sc i} and H$_2$ profiles (including the helium content) of our spirals using Eq.~\ref{eq:Frat_disc}. Some of the spirals, however, have gas surface density profiles with a more complex behavior. For these cases, following \cite{ceciVSFL}, we fit a `poly-exponential' profile of third degree
\begin{equation}
\label{eq:polyexp}
    \Sigma_{\rm HI,H_{2}}(R) = \Sigma_{\rm 0,pex}~e^{-R/R_{\rm pex}}~(1 + c_1 R + c_2 R^2 + c_3 R^3)~,
\end{equation}
with $\Sigma_{\rm,pex}, R_{\rm pex}, c_1, c_2, c_3$ the fitting parameters. The galaxies for which we use a poly-exponential fit are NGC~3621, NGC~4535, NGC~4736, and NGC~5033, in the case of H\,{\sc i}, and NGC~2403, NGC~3198, NGC~3351, NGC~4535, NGC~4725, NGC~5005, NGC~5033, and NGC~5055, in the case of H$_2$.

In a similar way as for the dwarfs, the gas discs of the spirals are also assumed to be thin in a preliminary step, but their final scale heights are obtained through our main fitting routine.

\subsubsection{Spirals: stellar bulge and disc}
As mentioned before, most of our galaxies have bulge-disc decomposition available from \citet{salo2015}. Those authors used a parametric modelling software (\textsc{galfit}, \citealt{galfit}) to decompose the 3.6~$\mu$m surface brightness profiles into different components. In a few cases we have neglected the contribution (less than 5 per cent of the total flux) from a PSF-like or bar component. In particular, we chose the parametrizations using an exponential disc plus a S\'ersic component. Some galaxies (NGC~3486, NGC~3675, NGC~4651, NGC~4736, NGC~5033, and NGC~5055) were modeled using two exponential discs to mimic a more complex behavior of their surface brightness profiles. We obtain the circular speed profile of our discs (up to the normalization $\Upsilon_{\rm d}$) with \textsc{galpynamics} using an exponential profile (Eq.~\ref{eq:exp_disc}) with parameters coming from \citealt{salo2015}. In the case of NGC~2403, NGC~3198, and NGC~3621, for which have no bulge-disc decomposition but the galaxies look bulgeless, we take the S4G 3.6~$\mu$m surfaces brightness profiles from \citet{munozmateos} and we fit two exponential discs (needed to reproduce the observed features in the profiles) following \cite{salo2015}. For all the galaxies with double stellar disc profiles we obtain their total $V_\ast$ by summing the individual contributions in quadrature. For the vertical distribution of all the stellar discs we make the same assumptions as for the stellar component of the dwarfs, assuming a sech$^2$ profile and a thickness $z_{\rm d} = 0.196 R_{\rm d}^{0.633}$ constant with radius. For the galaxies with two exponential discs we do not assume each has their own $z_{\rm d}(R_{\rm d})$ because usually one of the components has a low density and very large $R_{\rm d}$ (typically reproducing breaks/tails in surface brightness profiles). Instead, we combine both disc contributions and compute a half-light radius $R_{\rm eff}$, which is then transformed into a pseudo $R_{\rm d}$ using $R_{\rm eff} = 1.678 R_{\rm d}$; we then assume this pseudo $R_{\rm d}$ when computing $z_{\rm d} = 0.196 R_{\rm d}^{0.633}$. We note that the use of the expression  $R_{\rm eff} = 1.678 R_{\rm d}$ is not fully justified as it is only valid for exponential discs, but we consider it a reasonable mass-weighted compromise. In any case, this assumption does not alter our results below.\\

\begin{figure}
    \centering
    \includegraphics[scale=0.53]{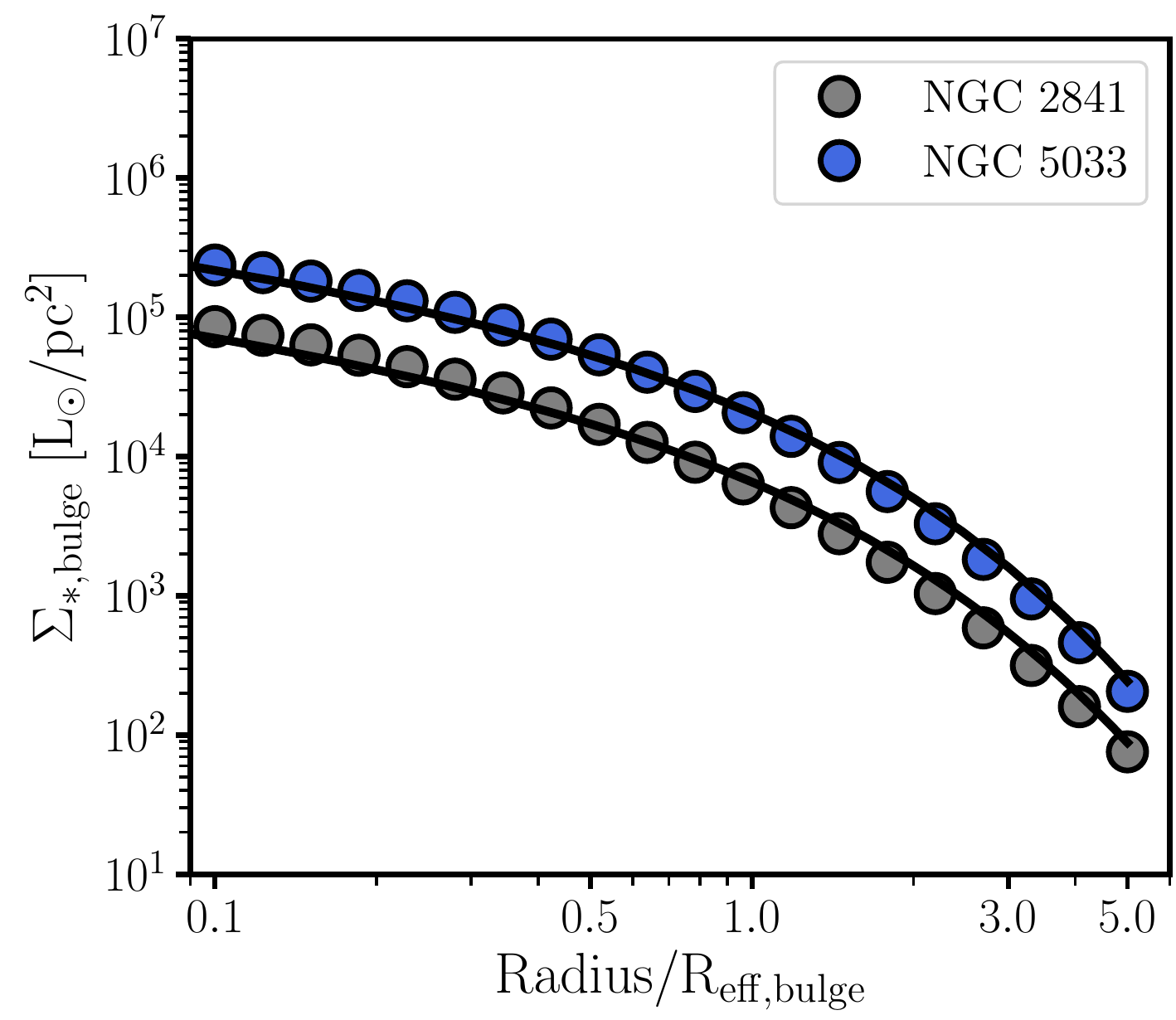}
    \caption{Comparison between the S\'ersic profiles of NGC~2841 (grey) and NGC~5033 (blue) as derived by \citet{salo2015}, and the projected surface density of our best-fitting spherical bulges for those galaxies (black curves). The x-axis is normalised to the effective radii of the bulges ($R_{\rm eff,bulge}$). We only show distances up to 5$R_{\rm eff,bulge}$ to ease the comparison, but a good agreement remains at all radii.}
    \label{fig:2bulges}
\end{figure}

\noindent
For the bulge components the situation is less straightforward. The S\'ersic profile describes surface densities with a flat (2D) geometry, while in practice bulges are spheroidal. Taking this into account is relevant because the force generated by a flattened distribution is different (stronger in the midplane) than that of an spheroid, and thus it can affect the recovery of the best-fitting dark matter halo parameters and the computation of the gas discs scale height. To obtain a realistic bulge component we assume that the bulges in our galaxies are spherical\footnote{We assume spherical symmetry for simplicity, but \textsc{galpynamics} can also deal with flattened spheroidal components.} and with a volume density described by a double power-law of the form:
\begin{equation}
\label{eq:bulge_alfabeta}
    \rho(r,z) = \dfrac{\rho_c} { (r/r_{\rm s})^\alpha (1 + (r/r_{\rm s}))^{\beta-\alpha}}~,
\end{equation}
with $r~=~\sqrt{R^2+z^2}$ the spherical radius, $\rho_{\rm c}$ and $r_{\rm s}$ a characteristic density and radius, respectively, and $\alpha$ and $\beta$ the power-law exponents. In practice, we obtain the best set ($\rho_{\rm c}$, $r_{\rm s}$, $\alpha$, $\beta$), using an MCMC routine based on the python package \texttt{emcee} \citep{emcee}, such that the resulting spherical bulge has a projected 2D surface density matching the S\'ersic profile reported by \cite{salo2015}. 
The resulting bulge is then given to \textsc{galpynamics} to obtain $V_{\rm b}$, while $\Upsilon_{\rm b}$ is obtained during our fitting routine. Fig.~\ref{fig:2bulges} shows the comparison between the S\'ersic profiles (circles) for the bulges of NGC~5033 and NGC~2814 and the surface density of the best-fitting spherical bulges we obtain (black curves). We find a similarly good agreement for our full sample.

\subsection{Deriving the thickness and flaring of the gas discs}
\label{sec:thickness}
Given an external mass distribution (e.g. a dark matter halo and a stellar disc), \textsc{galpynamics} is able to compute the gravitational potential of an additional galaxy component (e.g. a gas disc) and estimate its scale height under the assumption of vertical hydrostatic equilibrium. This is done iteratively taking into account the external gravitational potential and the self-gravity of the additional component through an iterative process \citep{iorio_phd}. For the interested reader, extensive and detailed discussion on the use of \textsc{galpynamics} to estimate scale heights can be found in \citet{iorio_phd,ceciVSFL,ceci_dwarfs}.

Effectively, \textsc{galpynamics} solves numerically the following expression (see \citealt{ceciVSFL,patra_H2}), which describes the 3D density profile $\rho(R,z)$ of the distribution whose scale height we aim to derive

\begin{equation}
\label{eq:rho_h}
    \rho(R,z) = \rho(R,0)\exp\left[ - \dfrac{\Phi(R,z) - \Phi(R,0)}{\sigma^2(R)} \right]~.
\end{equation}
In the above equation, $\rho(R,0)$ is the density profile of the gas (i.e. the radial profile of its volume density) evaluated in the disc midplane. In a similar way, $\Phi(R,0)$ is the total (dark matter plus stars plus gas disc(s)) gravitational potential $\Phi(R,z)$ evaluated in the midplane. Finally, $\sigma(R)$ is the radial profile of the gas velocity dispersion, assumed to be isotropic. Following \citet{ceciVSFL,ceci_dwarfs}, we fit the observed $\sigma(R)$ profiles with an exponential function $\sigma(R) = \sigma_0 e^{-R/R_\sigma}$, which provides a very good representation of the data and has the advantage of smoothing out small irregularities in the observed profiles, as shown in Fig.~\ref{fig:show_data}. The scale height $h(R)$ is defined as the standard deviation of Eq.~\ref{eq:rho_h}, and thus it can be obtained once $\Phi$ and $\sigma(R)$ are known. Note that some authors report the scale heights as the half width at half maximum of the vertical profiles, which should then be divided by a factor 1.777 if comparing with $h$, as we assume a Gaussian profile for the vertical distribution. In Sec.~\ref{sec:themethod} we explain how do we derive the thickness of the gas discs together with the best-fitting dark matter halo in a consistent way.


\subsection{Fitting the dark matter halo and mass-to-light ratios}
We assume that the dark matter haloes can be described by two different profiles depending on the galaxy mass. For the massive spirals we use a Navarro-Frenk-White (NFW, \citealt{nfw}) profile, while for the dwarfs we fit the so-called \textsc{core}NFW profile \citep{coreNFW,readAD}.

The density profile of the NFW profile is 
\begin{equation}
    \rho_{\rm NFW}(r) = \dfrac{4\rho_{\rm s}}{(r/r_{\rm s})(1 + r/r_{\rm s})^2}~,
\end{equation}
where $r~=~\sqrt{R^2+z^2}$, $r_{\rm s}$ is a scale radius, and $\rho_{\rm s}$ the volume density at $r_{\rm s}$. This volume density can be written as 
\begin{equation}
    \rho_{\rm s} = \dfrac{M_{200}}{16\pi r_{\rm s}^3 [ \ln(1+c_{200}) - c_{200}/(1+c_{200}) ] }~,
\end{equation}
with $M_{200}$ the dark matter mass within a radius $R_{200}$ where the average density is 200 times the critical density of the universe, and $c_{200} \equiv R_{200}/r_{\rm s}$ is the concentration; N-body simulations find a clear anti-correlation between $M_{200}$ and $c_{200}$ (e.g. \citealt{duttonmaccio2014,ludlow2014}).\\

\noindent
The \textsc{core}NFW profile allows the classical NFW profile to develop a core if this is needed to provide a better fit to slowly rising rotation curves. This is typically the case for dwarf galaxies and the well-known cusp-core problem (see \citealt{bullock2017} for a recent review).

The density profile of the \textsc{core}NFW halo is
\begin{equation}
    \rho_{\rm coreNFW}(r) = f^n \rho_{\rm NFW}(r) + \dfrac{nf^{n-1}(1-f^2)}{4\pi r^2 r_{\rm c}} M_{\rm NFW}(r)~,
\end{equation}
where $\rho_{\rm NFW}$ and $M_{\rm NFW}$ are the conventional density and mass for a NFW halo at a given radius, but $f~=~\tanh(r/r_{\rm c})$ is a function that generates a core of size $r_{\rm c} = 2.94~R_{\rm d}$. The parameter $n$ controls how strong is the transformation from cusp to core: $n=0$ means no core, recovering the NFW halo, and $n=1$ means a completely cored density profile. In practice, $n$ is defined as $n = \tanh(\kappa t_{\rm SF} / t_{\rm dyn})$, with $\kappa = 0.04$, $t_{\rm SF}$ the time whilst the galaxy has been forming stars (assumed to be 14~Gyr), and $t_{\rm dyn}$ the NFW dynamical time at the scale radius $r_{\rm s}$. For further details on all these parameters we refer the reader to \citet{coreNFW,readAD,read2017}. In the end, following the above prescriptions, the \textsc{core}NFW halo has the same two free parameters as the classic NFW halo: the mass $M_{200}$ and the concentration $c_{200}$. \\

\noindent
Together with the mass-to-light ratio for the stellar disc and the bulge, the dark matter halo parameters $M_{200}$ and $c_{200}$ are fitted using a MCMC routine based on \texttt{emcee}. Our scripts explores the parameter space looking for trial dark matter haloes and $\Upsilon$ to find the best set of values to minimise Eq.~\ref{eq:massmodel} using $\chi^2$ function.

For $\Upsilon_{\rm d}$, we use a Gaussian prior centered at $\Upsilon_{\rm d} = 0.5\ \rm{M_\odot/L_\odot}$, with a standard deviation of 0.11~dex, and with boundaries $0.1 < \Upsilon_{\rm d}/(\rm{M_\odot/L_\odot}) < 1.5$. Instead of fitting directly $\Upsilon_{\rm b}$, we define a parameter $f$ such that $\Upsilon_{\rm b} = f \Upsilon_{\rm d}$. We impose a Gaussian prior centered on $f=1.4$ with standard deviation of 0.1~dex and with boundaries $1 < f < 2$. These priors aim to break the degeneracy between disc, bulge, and halo in the centre of galaxies using values motivated by stellar population synthesis models (e.g. \citealt{mcgaugh_ML,meidt2014,querejeta2015}).


For the priors on $M_{200}$ and $c_{200}$, we investigate two different scenarios. In the first one, both $M_{200}$ and $c_{200}$ have flat and wide priors ($8 < \log(M_{200}/M_\odot) < 14$ and $2 < c_{200} < 50)$. In the second case, using the same boundary limits, we impose a Gaussian prior for c$_{200}$ which is centered on the $c_{200}-M_{200}$ relation of \citet{duttonmaccio2014}
\begin{equation}
    \log(c_{200}) = 0.905 - 0.101 \log\left(\dfrac{0.7 M_{200}}{10^{12} M_\odot}\right)~,
    \label{eq:c200M200}
\end{equation}
with a 1$\sigma$ standard deviation of 0.11~dex and assuming a Hubble parameter $h=0.7$. We note that our boundaries for the parameter space of $c_{200}$ are motivated by cosmological arguments (see \citealt{mcgaugh2003}). 

The approach detailed above allows us to check whether or not the $c_{200}-M_{200}$ is recovered from the data, or if it has to be imposed to obtain physically meaningful dark matter haloes. We discuss the results of these experiments in Section~\ref{sec:c200m200}.

\subsection{Self-consistent derivation of the thickness and dark matter halo}
\label{sec:themethod}
We now proceed to explain how we estimate the thickness of the gas discs and the best-fitting dark matter halo for each galaxy in our sample. This is done through an iterative processes based on \textsc{galpynamics} and our MCMC routine. The steps are as follows.

\begin{enumerate}
    \item [1.] Making use of \textsc{galpynamics} and assuming $\Upsilon_{\rm d} = \Upsilon_{\rm b}=1$, $V_{\rm d}$, $V_{\rm b}$ and the corresponding gravitational potentials $\Phi_{\rm d}$ and $\Phi_{\rm b}$ are derived from the stellar density profiles. 
    We also assume that the H\,{\sc i} and H$_2$ discs are razor-thin to obtain a first estimate of $\Phi_{\rm HI}$ and $\Phi_{\rm H_2}$ and their corresponding $V_{\rm HI}$ and $V_{\rm H_2}$.
    
    \item [2.] Using $V_{\rm d}$, $V_{\rm b}$, $V_{\rm HI}$, and $V_{\rm H_2}$, we fit $V_{\rm c}$ and we obtain a preliminary set ($\log(M_{200})$, $c_{200}$, $\Upsilon_{\rm d}$, $f$). Therefore, we can now update $\Phi_{\rm d}$ and $\Phi_{\rm b}$, and obtain $\Phi_{\rm DM}$. Traditional rotation curve decompositions would stop here.
    
    \item [3.] \textsc{galpynamics} computes the thickness of the H\,{\sc i} disc by solving numerically Eq.~\ref{eq:rho_h} while taking into account $\Phi_{\rm DM}$, $\Phi_{\rm d}$, $\Phi_{\rm b}$, $\Phi_{\rm H_2}$ (for a razor-thin H$_2$ disc) and the H\,{\sc i} disc self-gravity. This allow us to obtain a new estimate of $\Phi_{\rm HI}$.
    
    \item [4.] \textsc{galpynamics} computes the thickness of the H$_2$ disc taking into account $\Phi_{\rm DM}$, $\Phi_{\rm d}$, $\Phi_{\rm b}$, $\Phi_{\rm HI}$ (derived in the previous step), and the H$_2$ disc self-gravity.
    
    \item [5.] Step 3 and 4 are repeated iteratively until the changes in both scale heights, at all radii, are at least smaller than 10 per cent with respect to the previous iteration, which is a sensible choice given that the uncertainty in the scale height can be defined as the fractional error of the gas velocity dispersion ($\Delta h = h \Delta \sigma(R)/\sigma(R)$, see appendix E in \citealt{ceciVSFL}), for which 10 per cent is an appropriate uncertainty (see Fig.~\ref{fig:show_data}). Once these iterations converge, we end up with scale heights (and $\Phi$) for the H\,{\sc i} and H$_2$ discs that were derived taking into account each other plus the stellar disc, the bulge, and the dark matter halo.
    
    \item [6.] With all the new potentials, our MCMC routine finds a new best-fitting ($\log(M_{200})$, $c_{200}$, $\Upsilon_{\rm d}$, $f$) set, updating the mass model and generating a new set of potentials.
    
    \item [7.] Steps $3-6$ are repeated iteratively until ($\log(M_{200})$, $c_{200}$, $\Upsilon_{\rm d}$, $f$) converge. We adopt a convergence criterion such that the change in $M_{200}$, $c_{200}$, $\Upsilon_{\rm d}$, and $f$ between the last and penultimate iteration has to be less than 3 per cent. For most of our galaxies $2-3$ iterations are enough to reach convergence. 
\end{enumerate}

Due to our iterative process, the scale height of the gas discs, the disc and bulge stellar mass-to-light ratios, and the parameters of the dark matter halo, are derived in a fully consistent way taking into account the gravitational effects of each other.

\section{Results}
\label{sec:results}
\subsection{Mass models}

\begin{figure*}
    \centering
    \includegraphics[scale=0.55]{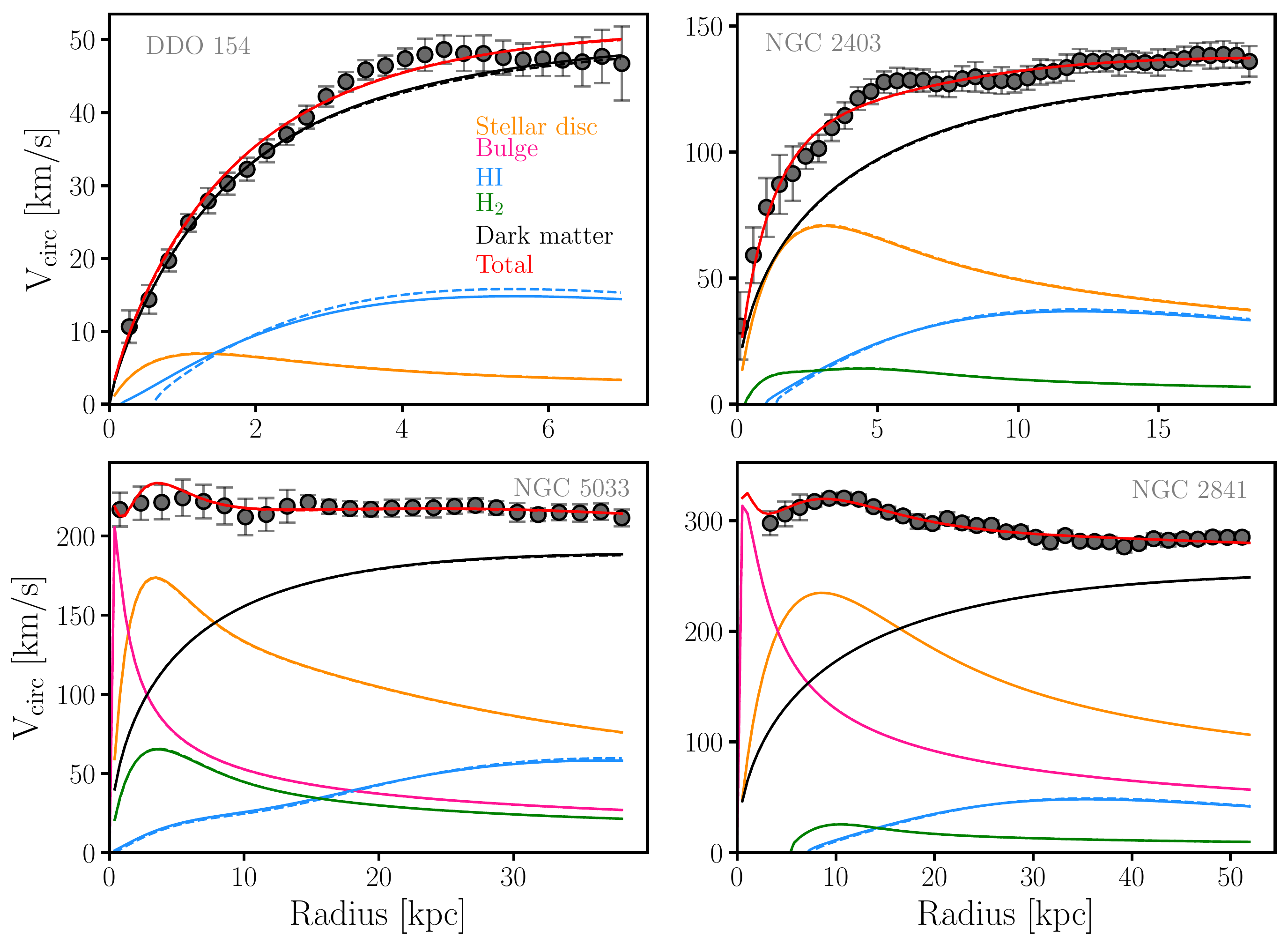}
    \caption{Mass models for the same representative galaxies as in Fig.~\ref{fig:show_data}. The models correspond to the case where $c_{200}$ is obtained with a prior on Eq.~\ref{eq:c200M200}. The points represent the circular speed profile of the galaxies (one out of each two points are shown for NGC~2403 and NGC~2841). The orange, pink, blue, green, and dark lines show the contribution to the circular speed by the stellar disc, bulge, H\,{\sc i} disc, H$_2$ disc, and dark matter halo, respectively. Dashed lines show the first iteration of our MCMC (with razor-thin gas discs), while solid lines show our final model (with flared gas discs). Similar plots for all our galaxies are available as supplementary material.}
    \label{fig:mmgau}
\end{figure*}

In Fig.~\ref{fig:mmgau}, we show the results of our rotation curve decomposition for the four representative galaxies displayed in Fig.~\ref{fig:show_data}. The figure corresponds to the case when $c_{200}$ was obtained using the prior on Eq.~\ref{eq:c200M200}. For each galaxy we show two mass models, one that assumes that the H\,{\sc i} and H$_2$ (if present) discs are razor-thin as in traditional methods, and one showing our final model after taking into account the flaring of the discs. Similar plots of the mass models (and corresponding posterior distributions) for our whole sample, for both types of priors on $c_{200}$, are available as \href{https://www.dropbox.com/sh/768phag6ttgym0j/AABZwerBgFofkkZ5npHW17YBa?dl=0}{supplementary material} to this manuscript. 

We obtain successful fits for all our sample when we use a prior on the $c_{200}-M_{200}$ relation: the best-fitting $V_{\rm c}$ faithfully reproduces the observations, and the posterior distributions of the best-fitting parameters are well behaved with a Gaussian or nearly Gaussian shape. When we use the flat prior on $c_{200}$ the fits are seemingly equally good, but for some galaxies (CVn I dwA, NGC~253, NGC~3486, NGC~3898, NGC~4535, NGC~5350, and UGC~8505) their $c_{200}$ values are not well constrained, as their posterior distributions go towards our lower bound $c_{200}=2$. In the following figures such galaxies are shown with different symbols to emphasise that their mass models are not fully reliable. As we discuss below, for the rest of the galaxies the mass models obtained under our two $c_{200}-$priors are in very good agreement with each other.

Fig.~\ref{fig:mmgau} shows that most of the time the overall effect of including the flaring of the gas discs is minor. This is not fully surprising: in the high-mass regime the potential of the galaxies in the inner regions is heavily dominated by the stellar components, where most of the gas mass resides. In the dwarf galaxies the gas dominates the baryonic content, but the dark matter haloes are more important from a dynamical point of view. 
In the case of DDO~154 (top left panel in Fig.~\ref{fig:mmgau}), one can appreciate that by not using the razor-thin disc assumption the `negative' velocities often seen in the circular speed of H\,{\sc i} discs (e.g. \citealt{deblok08,frank2016}) are gone. While in practice those velocities are considered negative (which is why the absolute values in Eq.~\ref{eq:massmodel} are needed), in the presence of H\,{\sc i} discs with a central depression, a test particle in the midplane would feel an outward acceleration due to the mass at larger radii. Because of this, the radial acceleration in the midplane is negative, and thus |$V_{\rm HI}$|$V_{\rm HI}$ is negative (see Eq.~\ref{eq:vc_potential}). Including the gas flaring strongly mitigates this effect: even when the central mass depression is there, the mass in the outer parts is distributed in a thicker disc, and thus a test particle in the midplane feels a weaker outwards acceleration, bringing |$V_{\rm HI}$|$V_{\rm HI}$ closer to zero and even to positive values. This is an interesting local effect only visible when considering the real geometry of the gas discs. 

\begin{figure}
    \centering
    \includegraphics[scale=0.5]{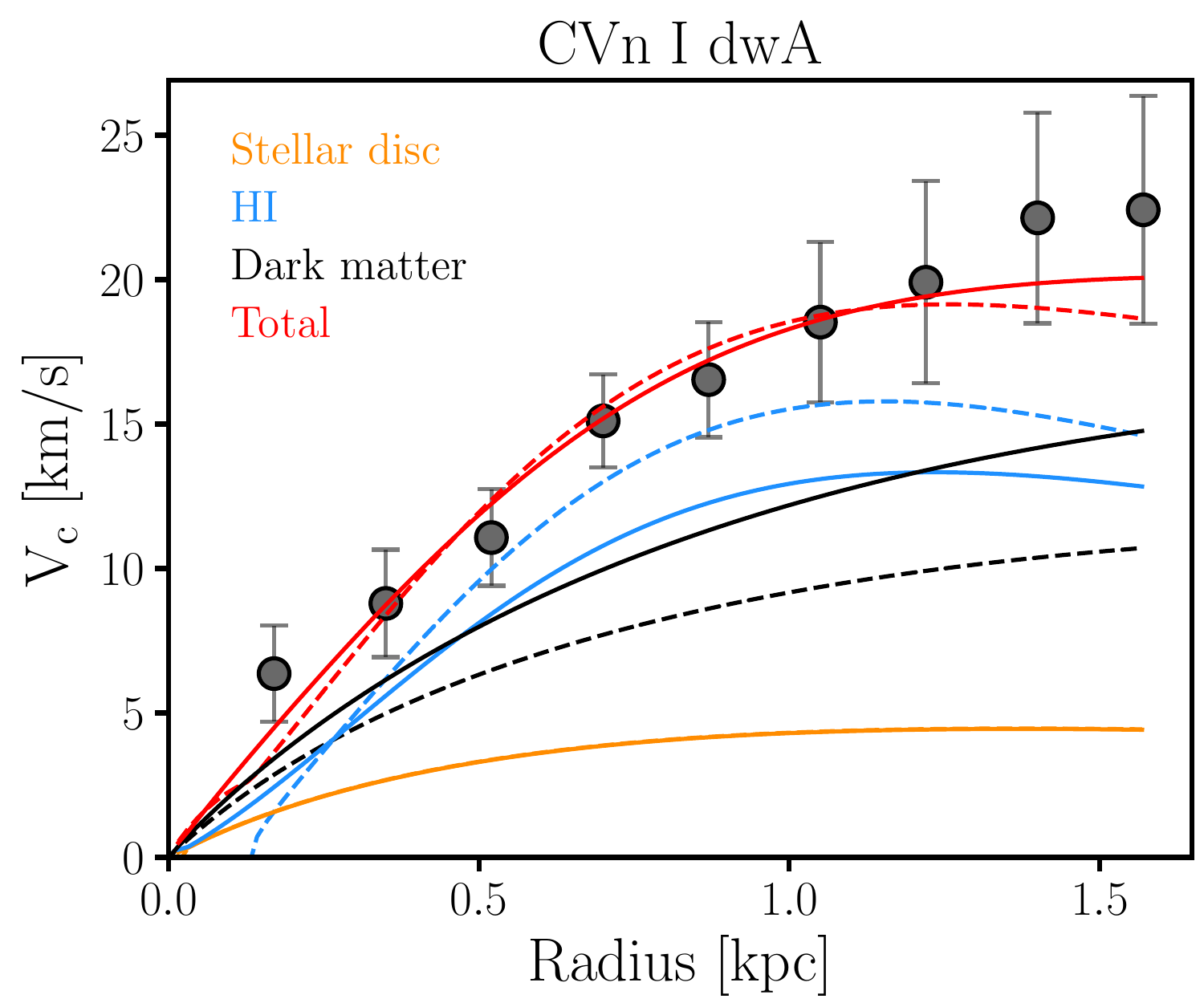}
    \caption{Mass model of the dwarf galaxy CVn~I~dwA. The solid (dashed) lines show the mass model when a flared (razor-thin) H\,{\sc i} disc is used. The large effect of the flaring on the mass model is clear.}
    \label{fig:cvidwa}
\end{figure}

In a few galaxies the effects of the flaring are important. In Fig.~\ref{fig:cvidwa} we show the mass model for the dwarf galaxy CVn~I~dwA, which shows the most extreme changes when including the H\,{\sc i} disc flaring within our sample. The flaring not only gets rid of the negative |$V_{\rm HI}$|$V_{\rm HI}$ in the innermost regions, but it also decreases the contribution of the gas disc at larger radii, allowing for a dark matter halo about three times more massive.\\

\noindent
We note here that our methodology described in Sec.~\ref{sec:themethod} does not take into account the uncertainty in the distances of our galaxy sample. While the distance could be included as a nuisance parameter in the MCMC fitting (e.g. with a Gaussian prior, see \citealt{li_massmodels,agc114905}), this is computationally prohibitive for our sample in our methodology as it would be required to compute the potential of each component on each iteration of the MCMC, and for each of these iterations to compute our scale heights. In practice, the distance to the galaxies tends to simply follow its prior and it only helps as a nuisance parameter to obtain appropriate uncertainties in the other quantities being fitted (see \citealt{li_massmodels}), we decided to follow a simplistic approach. We repeat our steps $1-7$ detailed in Sec.~\ref{sec:themethod}, but assuming the upper and lower $1\sigma$ uncertainties in the distance as reported in Table~\ref{tab:sample} (note that this also implies making the appropriate changes in the surface density profiles of the baryonic components). Then, for each galaxy we combine the three posterior distributions (one for each distance) of our parameters ($\log(M_{200}),c_{200},\Upsilon_{\rm d}, f$), and we re-obtain the 16$^{\rm th}$, 50$^{\rm th}$, and 84$^{\rm th}$ percentiles of the final total distributions, which we use to define the final ($\log(M_{200})$, $c_{200}$, $\Upsilon_{\rm d}$, $f$) set and its uncertainties. We report these values in Table~\ref{tab:results_gp} for the case imposing the $c_{200}-M_{200}$ relation as a prior, and in Table~\ref{tab:results_flat} for the flat prior.

\begin{table*}
\caption{Results of our rotation curve decomposition for the case when a prior on Eq.~\ref{eq:c200M200} is imposed. The columns give our fiducial values (the median of the posterior distributions) and their upper and lower uncertainties (corresponding to the $16^{\rm th}$ and $84^{\rm th}$ percentiles). Only galaxies with bulges have values of $f$. Note that all these quantities include the contribution from distance uncertainties.}
\label{tab:results_gp}
\begin{center} 
\begin{tabular}{lcccccccccccc}
	\hline \noalign{\smallskip}
	\multicolumn{1}{c}{Name} & \multicolumn{3}{c}{$\log(M_{200}/\rm{M_\odot})$} & \multicolumn{3}{c}{$c_{200}$} & \multicolumn{3}{c}{$\Upsilon_{\rm d} (\rm{M_\odot/L_\odot})$} & \multicolumn{3}{c}{$f$} \\ \noalign{\smallskip} 
 & $50^{\rm th}$ pctl & $\sigma-$ & $\sigma+$ & $50^{\rm th}$ pctl & $\sigma-$ & $\sigma+$ & $50^{\rm th}$ pctl & $\sigma-$ & $\sigma+$ & $50^{\rm th}$ pctl & $\sigma-$ & $\sigma+$  \\ \noalign{\smallskip} \hline
 CVn I dwA & 8.80 & $-$0.26 & + 0.26 & 18.08 & $-$4.57 & + 6.11 & 0.50 & $-$0.11 & + 0.11 & -- & -- & -- \\ 
DDO 52 & 10.36 & $-$0.16 & + 0.20 & 12.69 & $-$3.18 & + 4.02 & 0.50 & $-$0.11 & + 0.11 & -- & -- & -- \\ 
DDO 87 & 10.38 & $-$0.19 & + 0.36 & 12.66 & $-$3.74 & + 5.48 & 0.52 & $-$0.11 & + 0.11 & -- & -- & -- \\ 
DDO 126 & 9.95 & $-$0.14 & + 0.17 & 13.42 & $-$2.85 & + 3.29 & 0.46 & $-$0.11 & + 0.11 & -- & -- & -- \\ 
DDO 154 & 10.40 & $-$0.08 & + 0.09 & 9.23 & $-$0.95 & + 1.02 & 0.44 & $-$0.11 & + 0.11 & -- & -- & -- \\ 
DDO 168 & 10.56 & $-$0.16 & + 0.20 & 12.37 & $-$2.64 & + 3.01 & 0.43 & $-$0.11 & + 0.11 & -- & -- & -- \\ 
DDO 210 & 8.35 & $-$0.57 & + 0.57 & 20.02 & $-$5.42 & + 7.30 & 0.50 & $-$0.11 & + 0.11 & -- & -- & -- \\ 
NGC 0253 & 12.31 & $-$0.23 & + 0.30 & 6.60 & $-$1.61 & + 2.01 & 0.32 & $-$0.03 & + 0.03 & 1.40 & $-$0.10 & + 0.10 \\ 
NGC 1313 & 11.91 & $-$0.12 & + 0.16 & 8.51 & $-$1.41 & + 1.47 & 0.30 & $-$0.08 & + 0.08 & 1.38 & $-$0.10 & + 0.10 \\ 
NGC 2366 & 10.47 & $-$0.09 & + 0.11 & 15.19 & $-$2.97 & + 3.30 & 0.50 & $-$0.11 & + 0.11 & -- & -- & -- \\ 
NGC 2403 & 11.60 & $-$0.05 & + 0.06 & 11.07 & $-$1.40 & + 1.54 & 0.42 & $-$0.07 & + 0.07 & -- & -- & -- \\ 
NGC 2841 & 12.52 & $-$0.05 & + 0.05 & 8.81 & $-$1.28 & + 1.69 & 0.84 & $-$0.08 & + 0.08 & 1.37 & $-$0.09 & + 0.09 \\ 
NGC 3198 & 11.62 & $-$0.02 & + 0.03 & 12.24 & $-$1.36 & + 1.47 & 0.40 & $-$0.07 & + 0.07 & -- & -- & -- \\ 
NGC 3351 & 11.78 & $-$0.13 & + 0.16 & 8.22 & $-$1.88 & + 2.36 & 0.56 & $-$0.05 & + 0.05 & 1.41 & $-$0.09 & + 0.09 \\ 
NGC 3486 & 12.30 & $-$0.15 & + 0.18 & 4.60 & $-$0.89 & + 1.11 & 0.35 & $-$0.08 & + 0.10 & 1.41 & $-$0.10 & + 0.10 \\ 
NGC 3621 & 11.62 & $-$0.04 & + 0.05 & 11.80 & $-$1.38 & + 1.50 & 0.22 & $-$0.05 & + 0.05 & -- & -- & -- \\ 
NGC 3675 & 12.25 & $-$0.13 & + 0.15 & 8.66 & $-$1.86 & + 2.43 & 0.46 & $-$0.10 & + 0.14 & -- & -- & -- \\ 
NGC 3898 & 12.85 & $-$0.15 & + 0.17 & 4.76 & $-$1.11 & + 1.89 & 0.63 & $-$0.11 & + 0.19 & 1.42 & $-$0.10 & + 0.11 \\ 
NGC 3992 & 11.83 & $-$0.05 & + 0.05 & 20.55 & $-$6.41 & + 12.18 & 0.76 & $-$0.10 & + 0.10 & 1.41 & $-$0.10 & + 0.10 \\ 
NGC 4535 & 12.62 & $-$0.20 & + 0.26 & 6.62 & $-$1.61 & + 1.93 & 0.52 & $-$0.08 & + 0.08 & 1.43 & $-$0.10 & + 0.10 \\ 
NGC 4536 & 11.99 & $-$0.14 & + 0.17 & 7.43 & $-$1.47 & + 1.66 & 0.27 & $-$0.04 & + 0.05 & 1.37 & $-$0.10 & + 0.10 \\ 
NGC 4559 & 11.41 & $-$0.06 & + 0.07 & 9.63 & $-$1.58 & + 1.81 & 0.36 & $-$0.08 & + 0.08 & 1.39 & $-$0.10 & + 0.10 \\ 
NGC 4651 & 11.61 & $-$0.09 & + 0.09 & 17.89 & $-$4.53 & + 7.47 & 0.62 & $-$0.13 & + 0.16 & 1.43 & $-$0.10 & + 0.10 \\ 
NGC 4698 & 12.17 & $-$0.09 & + 0.10 & 10.07 & $-$2.08 & + 2.70 & 0.50 & $-$0.09 & + 0.12 & 1.38 & $-$0.10 & + 0.10 \\ 
NGC 4725 & 11.84 & $-$0.06 & + 0.08 & 12.65 & $-$2.76 & + 3.36 & 0.67 & $-$0.08 & + 0.08 & 1.40 & $-$0.10 & + 0.10 \\ 
NGC 4736 & 11.33 & $-$0.09 & + 0.12 & 17.97 & $-$3.37 & + 3.72 & 0.21 & $-$0.03 & + 0.03 & 1.28 & $-$0.10 & + 0.10 \\ 
NGC 5005 & 12.21 & $-$0.24 & + 0.32 & 9.01 & $-$2.51 & + 3.24 & 0.50 & $-$0.06 & + 0.08 & 1.36 & $-$0.09 & + 0.09 \\ 
NGC 5033 & 12.07 & $-$0.05 & + 0.05 & 11.25 & $-$1.41 & + 1.60 & 0.31 & $-$0.05 & + 0.07 & 1.40 & $-$0.09 & + 0.10 \\ 
NGC 5055 & 12.01 & $-$0.04 & + 0.04 & 9.14 & $-$1.02 & + 1.14 & 0.37 & $-$0.04 & + 0.04 & -- & -- & -- \\ 
NGC 5350 & 12.22 & $-$0.11 & + 0.13 & 5.05 & $-$1.03 & + 1.27 & 0.47 & $-$0.08 & + 0.10 & 1.46 & $-$0.10 & + 0.10 \\ 
UGC 8508 & 10.43 & $-$0.39 & + 0.43 & 12.64 & $-$3.60 & + 5.23 & 0.50 & $-$0.11 & + 0.11 & -- & -- & -- \\ 
WLM & 10.33 & $-$0.22 & + 0.24 & 8.83 & $-$1.80 & + 2.34 & 0.50 & $-$0.11 & + 0.11 & -- & -- & -- \\ 
   \hline
    \end{tabular}
  \end{center}
\end{table*}

\begin{figure}
    \centering
    \includegraphics[scale=0.4]{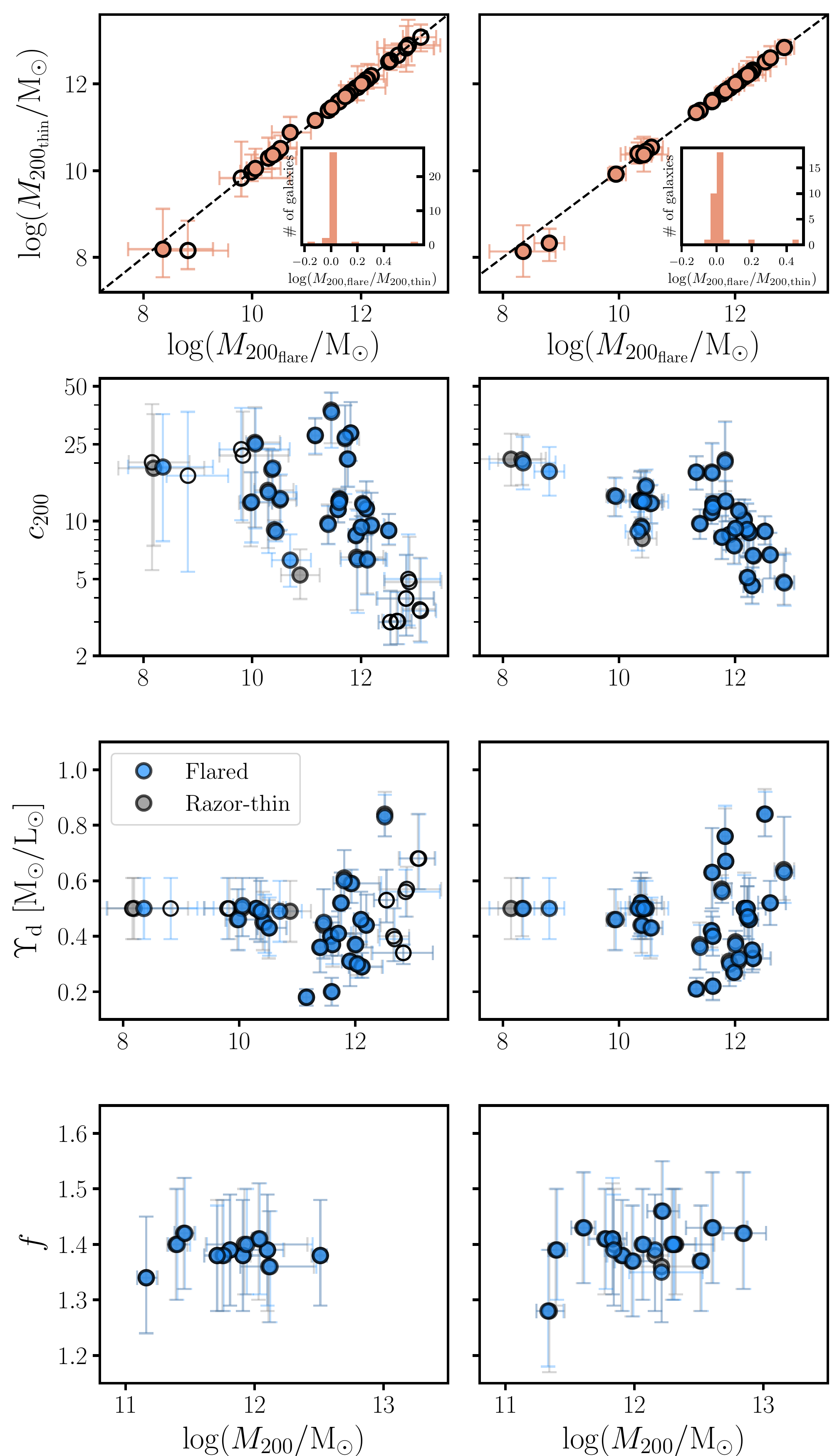}
    \caption{Comparison between the recovered parameters from our mass modelling when the gas discs are considered razor-thin (gray circles) and when the flaring is considered (blue circles); most of the time the values overlap. Empty symbols correspond to those galaxies for which we do not obtain well-constrained dark matter halo parameters when using a flat prior on $c_{200}$. Panels on the \textit{left} concern the results obtained with a flat prior on $c_{200}$, while the \textit{right} panels correspond to the results obtained assuming the prior on Eq.~\ref{eq:c200M200}. The \textit{top} panels show a direct comparison so there is no distinction between blue and grey circles; the dashed line indicates the 1:1 relation, while the insets show the distribution of the differences between $M_{200}$ when derived with the flared and razor-thin gas discs.}
    \label{fig:comparison}
\end{figure}

In Fig.~\ref{fig:comparison} the effects of the gas flaring can be seen more clearly, as we show the comparison between the values of $M_{200}$, $c_{200}$, $\Upsilon_{\rm d}$, and $f$ recovered using razor-thin and flared gas discs, as a function of $M_{200}$. The left panels show the comparison for our uninformative prior on $c_{200}$ while the right panel shows it for the prior on $c_{200}$ following the $c_{200}-M_{200}$ relation of \cite{duttonmaccio2014}. The parameters in the figure are obtained after considering the distance uncertainty as described in the previous paragraph.

Regarding $M_{200}$, the values derived with both razor-thin and flared discs are compatible with the 1:1 relation. As shown in the inset, the differences between the two derivations are close to zero. The two galaxies with a difference larger than $+0.2$~dex are dwarfs (the most extreme one being CVn~I~dwA; Fig.~\ref{fig:cvidwa}), where the flaring is expected to be more important as H\,{\sc i} dominates the baryonic budget of the galaxies. No systematic trends as a function of $M_{200}$ are clearly visible in the data, but this should be studied more carefully with larger samples of low-mass galaxies. 

The second row of panels in Fig.~\ref{fig:comparison} shows the the recovered concentration parameters. Again, within the uncertainties the razor-thin- and flared-derived values are compatible. When using a flat prior, the concentration of some galaxies at high $M_{200}$ tend towards $c_{200}=2$, the lower limit of our priors. However, those galaxies also have low $c_{200}$ values when imposing the prior on Eq.~\ref{eq:c200M200}, and in general the values obtained using both priors are in agreement with each other. In Section~\ref{sec:c200m200} we discuss in detail the dark matter concentration-halo mass relation.

The third row shows the disc mass-to-light ratios. Motivated by stellar population synthesis models, we used a Gaussian prior centered at $\Upsilon_{\rm d} = 0.5\pm0.1\ \rm{M_\odot/L_\odot}$. Interestingly, the values of $\Upsilon_{\rm d}$ show significant spread (which does not correlate with optical colours\footnote{We checked this by recovering the ($B-V$) colours of our sample from the HyperLeda database \citep{HyperLeda}.}) at $M_{200} \gtrsim 10^{11}~\rm{ M_\odot}$ (i.e. for all the massive disc galaxies). At these masses, the median value for our flat (Eq.~\ref{eq:c200M200}) prior scenario is is $0.41\ (0.46)\ \rm{M_\odot/L_\odot}$, with a standard deviation of $0.15\ (0.17)\ \rm{M_\odot/L_\odot}$ and values in the range $0.18\ (0.21)-0.83\ (0.84)\ \rm{M_\odot/L_\odot}$. We checked with our two most extreme galaxies in terms of $\Upsilon_{\rm d}$ (NGC~2841 with $\Upsilon_{\rm d}=0.83\ \rm{M_\odot/L_\odot}$ and NGC~4736 with $\Upsilon_{\rm d}=0.18\ \rm{M_\odot/L_\odot}$) the consequences of assuming a value of $\Upsilon_{\rm d}=0.5\ \rm{M_\odot/L_\odot}$ often assumed when studying scaling relations (e.g. \citealt{lelliBTFR,paperIBFR}). In the case of NGC~2841 one obtains a similar $M_{200}$ but the fit to the rotation curve is slightly worse than in Fig.~\ref{fig:mmgau}; moreover, the concentration of the galaxy moves away from the $c_{200}-M_{200}$ relation (see below). For NGC~4736 the mass model always overestimates the observed circular speed. This shows that the value of $\Upsilon_{\rm d} = 0.5\ \rm{M_\odot/L_\odot}$ is representative, but for some galaxies in our sample it would bias the estimation of the stellar mass (and concentration) by a significant amount (see also the discussion in \citealt{anastasia_SED} on estimating $\Upsilon_{\rm d}$ with different methods). For smaller galaxies, Fig.~\ref{fig:comparison} may give the impression that $\Upsilon_{\rm d} = 0.5\ \rm{M_\odot/L_\odot}$ is perfectly appropriate. However, it should be noted that their posterior distributions on $\Upsilon_{\rm d}$ simply follow their Gaussian priors. This is because the contribution of $V_{\rm d}$ to the mass models of dwarfs is small, and thus our MCMC fitting loses constraining power on $\Upsilon_{\rm d}$; because of this, it is likely that the real uncertainties in the $\Upsilon_{\rm d}$ for the dwarfs are underestimated by $20-40$ per cent. The effects of the gas discs flaring are negligible (up to the uncertainties just described for the dwarf galaxies), as seen from the overlap of grey and blue circles in the third panels of Fig.~\ref{fig:comparison}.

Finally, the bottom panels of Fig.~\ref{fig:comparison} show the values of $f$ as a function of $M_{\rm 200}$; note the different range in the x-axis with respect to the other panels, as $f$ is only a parameter for the massive galaxies with bulge. There are no visible differences between values derived with razor-thin or flared gas discs, and within the uncertainties the prior of $f = 1.4\pm0.1$ provides satisfactory fits to the data.

\subsection{Scale heights}
\begin{figure*}
    \centering
    \includegraphics[scale=0.51]{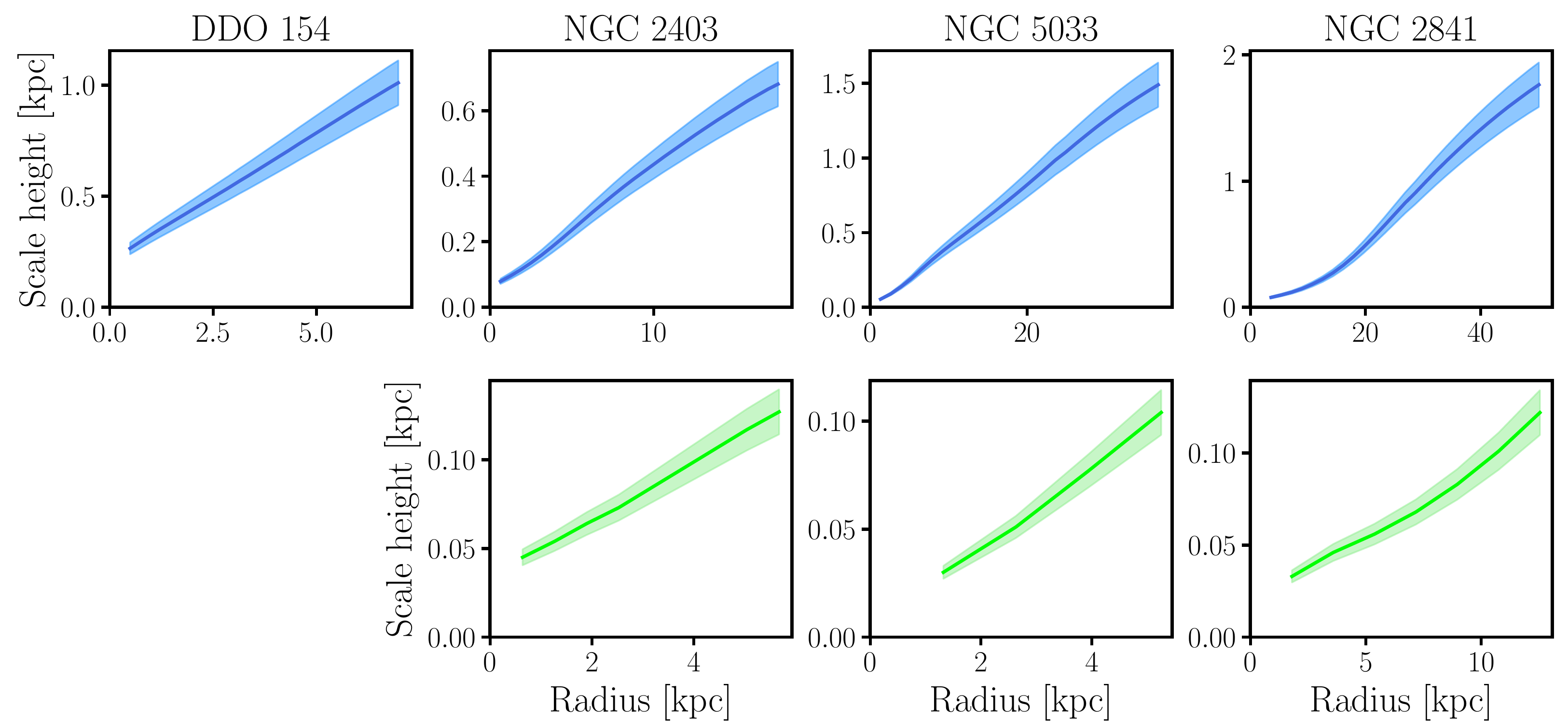}
    \caption{Example of H\,{\sc i} (\textit{top}) and H$_2$ (\textit{bottom}) scale heights in our sample. The solid blue and green lines show our estimations, while the colour bands show the assumed uncertainty of 10 per cent, motivated by the uncertainty in the gas velocity dispersion (see Fig.~\ref{fig:show_data}). We only show the scale heights for our mass models with a prior on the $c_{200}-M_{200}$ relation, but the profiles obtained with the flat prior largely overlap.} 
    \label{fig:example_h}
\end{figure*}

\begin{figure}
    \centering
    \includegraphics[scale=0.5]{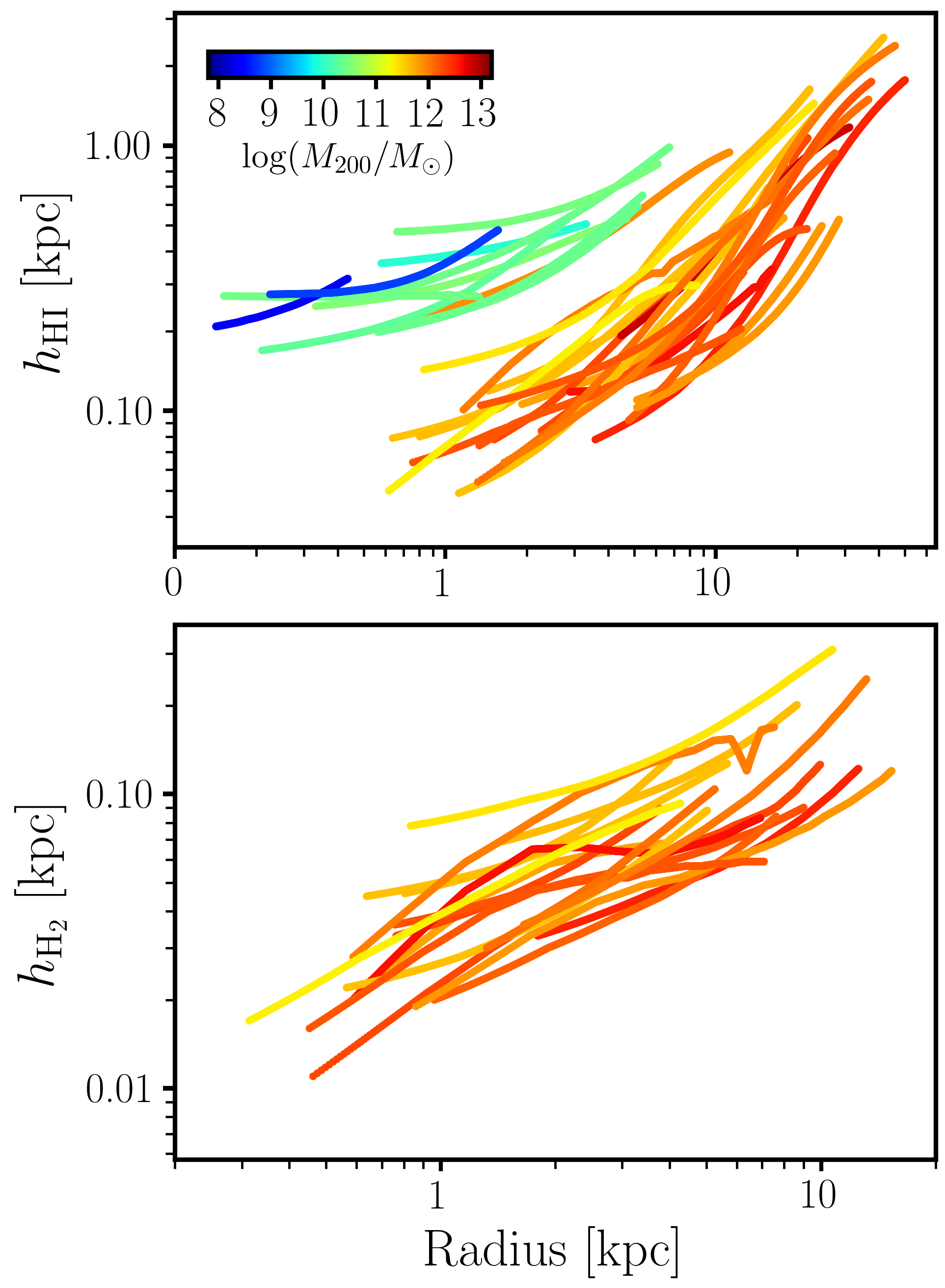}
    \caption{H\,{\sc i} (\textit{top}) and H$_2$ (\textit{bottom}) scale heights for our galaxy sample. Galaxies are colour-coded according to their halo mass.}
    \label{fig:all_h}
\end{figure}

Fig.~\ref{fig:example_h} shows the H\,{\sc i} (blue line and band) and H$_2$ (green line and band) scale heights for the same four representative galaxies as before. Fig.~\ref{fig:all_h} shows the scale heights for our full sample, and tables giving the scale heights are available \href{{https://www.dropbox.com/sh/768phag6ttgym0j/AABZwerBgFofkkZ5npHW17YBa?dl=0}}{on-line}. Naturally, the scale heights increase with increasing galactocentric distance. At all radii the scale height of the H\,{\sc i} discs is larger than for the H$_2$ discs. This is expected given that the velocity dispersion of H\,{\sc i} is higher. 

While the overall behavior of the flaring is the same in all our galaxies, the steepness and normalization change from galaxy to galaxy (see Fig.~\ref{fig:all_h}). Previous observational works have postulated a `universality' of the H\,{\sc i} and H$_2$ scale heights in late-type galaxies when normalised to characteristic scales \citep{patra_H2,patra_HIdwarfs,patra_HIspirals}. In particular, using a sample of disc galaxies \citet{patra_H2,patra_HIspirals} with $R$ normalised to units of $R_{25}$ (the radius at which a $B-$band isophote reaches 25 mag/arcsec$^2$), and $h(R)$ to be unity at a radius of $0.3 R_{25}$. \citet{patra_HIdwarfs} instead adopted a pseudo-isothermal parameterisation to fit to a set of rotation curves of dwarf galaxies, and used the `core radius' to normalise $R$. The normalised $h(R)$ values seem to have a linear dependency on the normalised radial scale, although the scatter is fairly large. The logic behind using $R_{25}$ or the core radius is that the claimed universal shape of the $h(R)$ profiles is a dynamical effect. Thus, it is appropriate to re-scale using characteristic scales related to the radii at which the stellar or dark matter component are dominant. On the other hand using $0.3 R_{25}$ as a normalization factor of $h(R)$ seems less justified.


In Fig.~\ref{fig:all_h} we plot the scale heights for our galaxy sample colour-coded by the halo mass. There is a general trend of low-mass galaxies having thicker discs than their more massive counterparts, as expected given their weaker potentials. At fixed $M_{200}$ the profiles are somewhat homogeneous, although the scatter is always large. We explore introducing the above normalizations as well as more dynamically-motivated values such as normalizing the radius to a fraction of the virial radius and the re-scale $h(R)$ according to the virial velocities, but significant scatter remains. We thus find little evidence for a universal $h(R)$ profile. This is not fully surprising as $h(R)$ depends strongly on the gas velocity dispersion, the dark matter halo, the stellar mass-to-light ratio,  and the bulge-to-total mass ratio, which show a large spread in our sample. While there are a number of significant differences between the derivation of $h(R)$ by \cite{patra_H2,patra_HIdwarfs,patra_HIspirals} and ours (see for instance the discussion in \citealt{ceci_dwarfs}), it is likely that the different conclusions arise mostly from our larger sample with a wider variety in gas kinematics, halo mass, bulge ratio, and mass-to-light ratio. 

\section{Implications and discussion}
\label{sec:discussion}
Our data and the parameters we recover from our fitting procedure allow us to revisit some of the most important scaling relations concerning the connection between galaxies and their dark matter haloes. In this section, we discuss three of them: the stellar--to--halo mass relation, the baryon--to--halo mass relation, and the halo concentration--mass relation.

\subsection{Stellar--to--halo mass relation}
The stellar--to--halo mass relation (SHMR) has been widely studied in the context of understanding the star formation efficiency as a function of the halo mass (e.g. \citealt{moster2013,aldo2015,wechsler18,posti_galaxyhalo,romeo2020}). The classical picture based on abundance-matching techniques is that the SHMR increases monotonically from the dwarf regime till $M_\ast \approx 5\times10^{11}~\rm{ M_\odot}$, where it flattens. The relation, however, might be different if galaxies are split in quenched and star-forming \citep{aldo2015,posti_gcs}. 

To compute $M_\ast$ for our sample, we integrate the luminosity profiles of the bulge and disc components up to $R_{200}$. We do this rather than integrating to infinity in order to measure the mass of the different baryonic and dark components within the same observed radius. In any case, this integration limit has very little impact in the integrated stellar mass as the luminosity profiles have values close to zero at such large distances, and it has no repercussion on our results. Once the total luminosity of disc and bulge is estimated, we multiply by the corresponding $\Upsilon_{\rm d}$ and by $f\Upsilon_{\rm d}$ in the case of the bulge, and we add the mass of both components. The uncertainties in all the quantities are estimated using Monte Carlo realizations, including distance and best-fitting parameters uncertainties. The mass of the baryonic components of our galaxies are listed in Table~\ref{tab:masses}.

\begin{figure}
    \centering
    \includegraphics[scale=0.5]{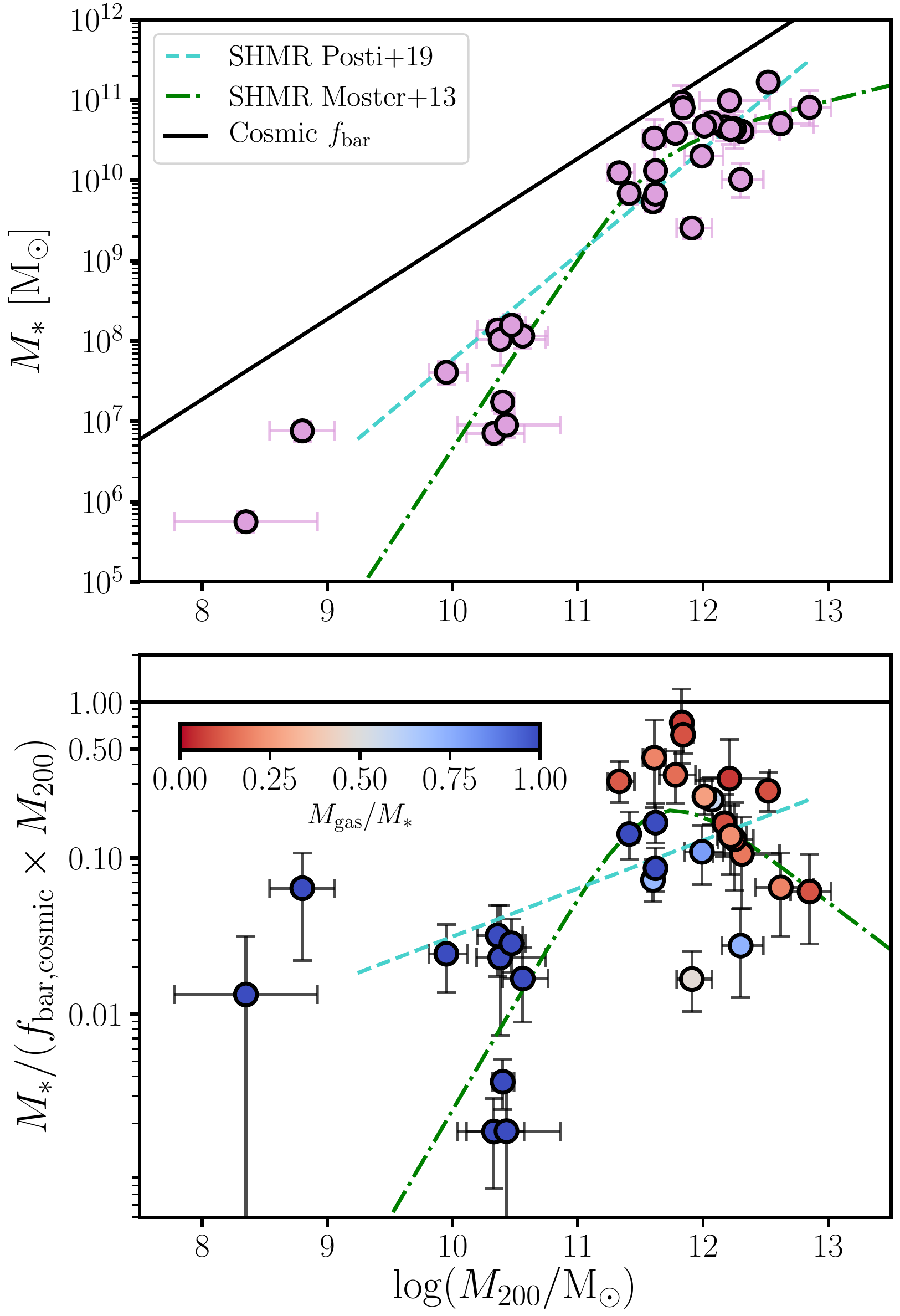}
    \caption{\textit{Top:} Stellar-to-halo mass relation for our galaxy sample. \textit{Bottom:} $M_\ast/M_{200}$ ratio normalised to the cosmological baryon fraction vs. $M_{200}$. Galaxies are colour-coded according to their stellar-to-cold gas content $M_{\rm gas}/M_{\rm \ast}$. In both panels the black solid line shows the cosmological baryon fraction $f_{\rm bar,cosmic}$. For comparison, we plot with a green semi-dashed curve the SHMR relation from \citet{moster2013}, while the teal dashed line shows a fit to the analysis by \citet{postinomissing}.}
    \label{fig:mstarm200}
\end{figure}

In the top panel of Fig.~\ref{fig:mstarm200} we show the SHMR for our galaxy sample, using the values derived with our prior on the $c_{200}-M_{200}$ relation. The black solid lines show the cosmic limit, given by $M_{200}/f_{\rm bar,cosmic}$. We can see that $M_\ast$ increases with $M_{200}$ rather linearly (in log-log space), although the scatter at high $M_{200}$ is large. We compare our measurements with the abundance matching-based SHMR from \citet{moster2013}, and with a linear fit to the data from \cite{postinomissing}, who performed the rotation curve decomposition of galaxies in the SPARC data base \citep{sparc}. Our data points appear to scatter around both SHMRs, but our low-number statistics prevent us from drawing conclusions on this, albeit the scatter is likely real as it is also present when studying larger samples \citep{posti_gcs}.


In the bottom panel of Fig.~\ref{fig:mstarm200} we show the $M_\ast/M_{200}$ ratio normalised to the average cosmological baryon fraction ($f_{\rm bar,cosmic} = \Omega_{\rm b}/\Omega_{\rm m} \approx 0.187$, \citealt{planck2020_fbar}). We can see that at fixed $M_{200}$ galaxies with lower $M_{\rm gas}/M_{\rm \ast}$ ratios have higher $M_\ast/M_{200}$ values. In addition to this, the disc galaxies with the largest $M_{200}$ have $M_\ast/M_{200}$ as low as dwarf galaxies, despite their much lower $M_{\rm gas}/M_{\rm \ast}$. We can see from both panels that at the high-mass regime four spiral galaxies have baryon fractions between $\sim 50-100$ percent of $f_{\rm bar,cosmic}$. The existence of galaxies with nearly all of their baryons in the form of stars at $M_\ast \sim 10^{11}~\rm{ M_\odot}$ was in fact suggested recently by \citet{postinomissing}.

\begin{figure}
    \centering
    \includegraphics[scale=0.45]{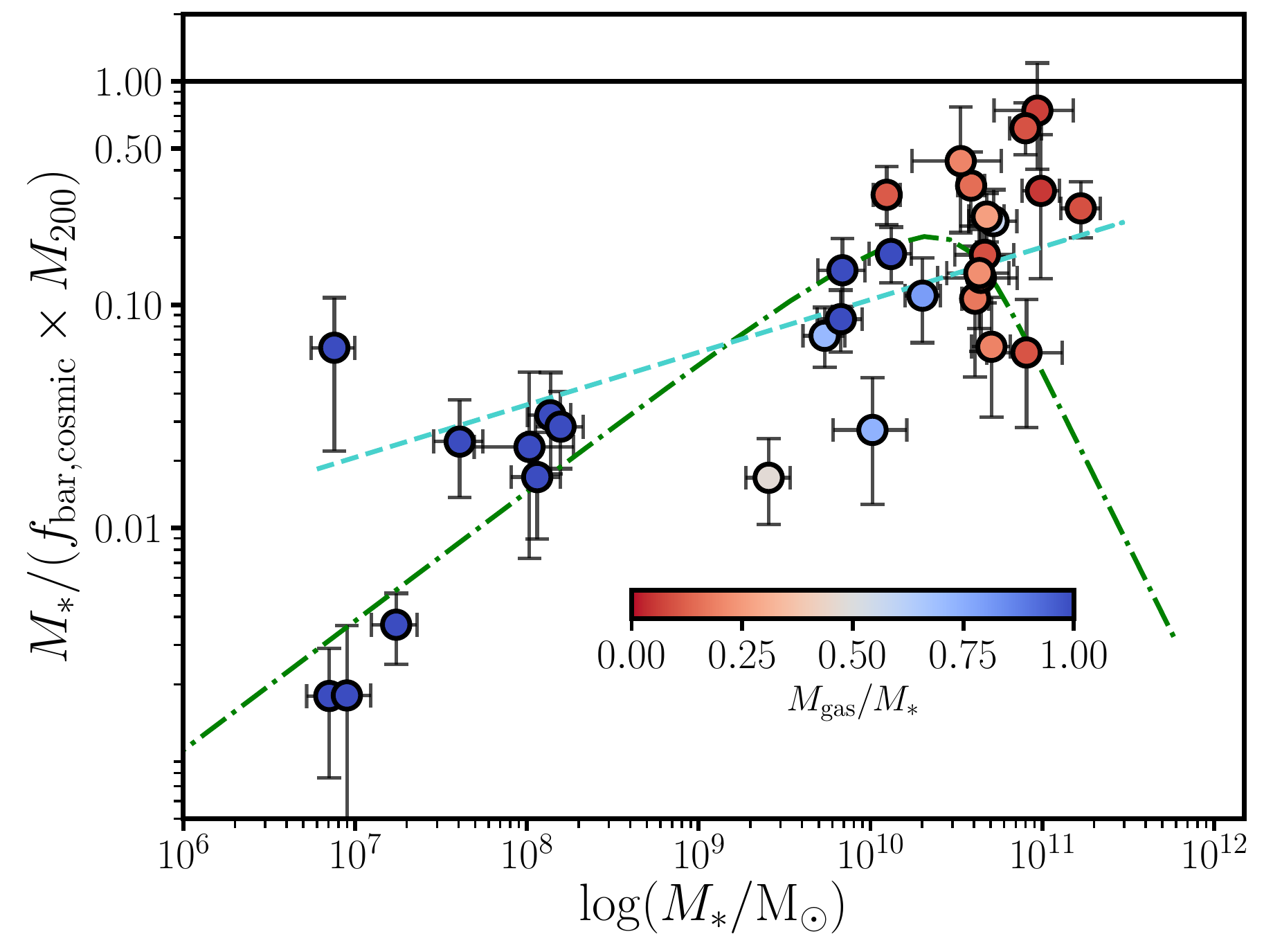}
    \caption{$M_\ast/M_{200}$ ratio normalised to the cosmological baryon fraction vs. $M_\ast$. Colours and curves are as in Fig.~\ref{fig:mstarm200}.}
    \label{fig:fstarmstar}
\end{figure}

To see this more clearly, in Fig.~\ref{fig:fstarmstar} we plot the stellar mass fraction now as a function of $M_\ast$. This shows that the galaxies with the highest stellar masses have also the highest $M_\ast/M_{200}$ ratios. This is in qualitative agreement with the results from \cite{postinomissing,posti_galaxyhalo,posti_gcs}, and deviates significantly from the abundance-matching predictions.

\subsection{Baryon--to--halo mass relation}
We now focus on the baryon--to--halo mass relation (BHMR). Arguably, the BHMR is more fundamental than the SHMR, as it incorporates simultaneously the stellar and gas content. It is also more robust, especially at the low-mass regime since dwarf irregular galaxies are gas-dominated.

To compute the baryonic mass we consider the stellar mass as defined before, together with the neutral and molecular gas content. Similarly as for $M_\ast$, we compute $M_{\rm HI}$ and $M_{\rm H_2}$ by integrating the functional forms fitted to the observed surface densities, from 0 to $R_{200}$.
Strictly speaking, our $M_{\rm bar}$ is not the total baryonic mass as it neglects the mass of the warm and hot ionised gas. This assumption is commonly made given that the warm and hot ionised gas are not expected to contribute significantly to the whole baryonic budget near galaxy discs, and because it is extremely challenging to obtain reliable mass measurements of these gas phases. Under the same logic, our $M_{\rm gas}$ values should be seen as the mass of the cold ISM gas.


\begin{figure}
    \centering
    \includegraphics[scale=0.5]{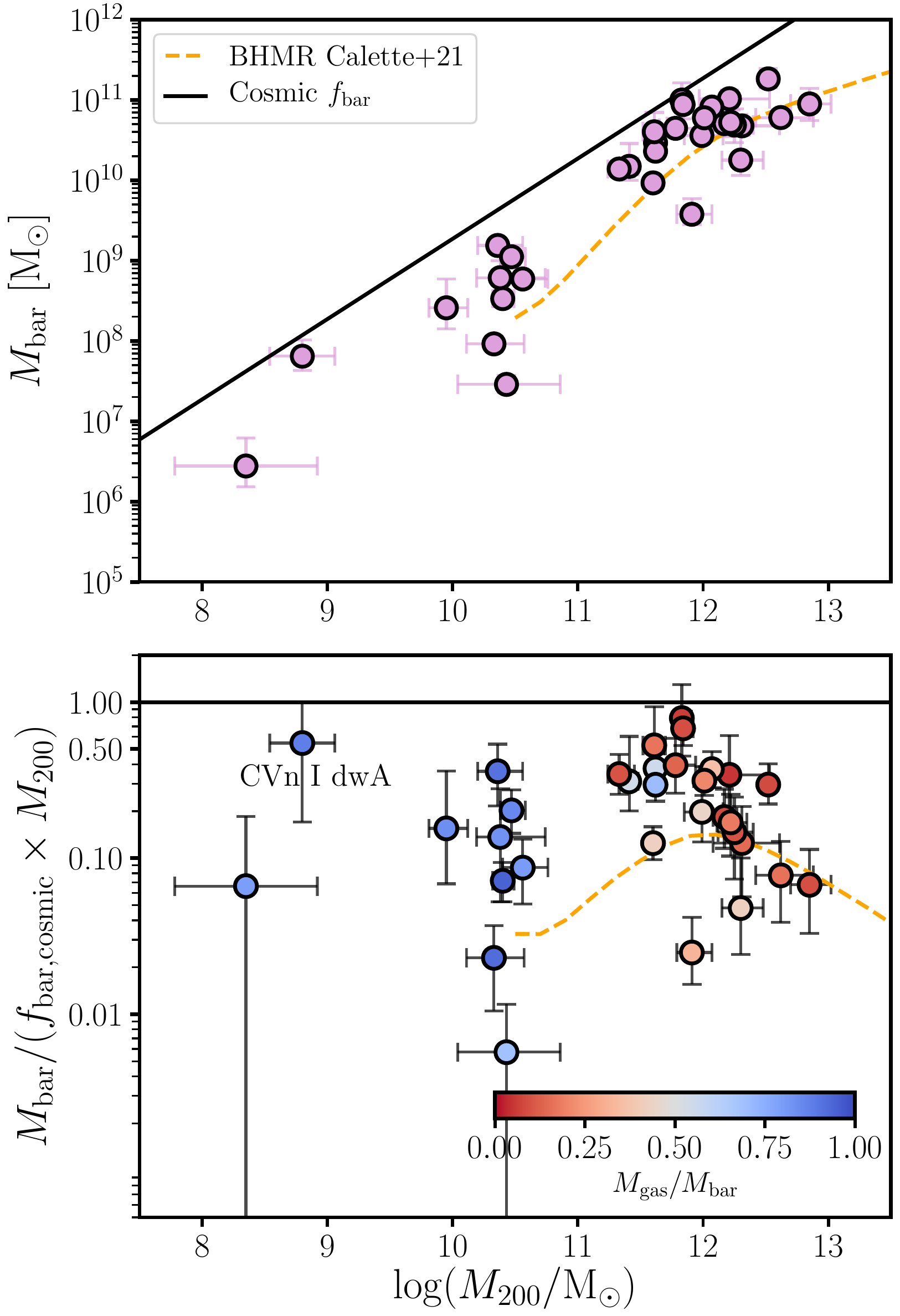}
    \caption{\textit{Top:} Baryon--to--halo mass relation for our galaxy sample. \textit{Bottom:} $M_{\rm bar}/M_{200}$ ratio normalised to the cosmological baryon fraction vs. $M_{200}$. Galaxies are colour-coded according to their stellar-to-cold gas content $M_{\rm gas}/M_{\rm \ast}$. In both panels the black solid line shows the cosmological baryon fraction $f_{\rm bar,cosmic}$, and the orange dashed curve shows the semi-empirical relation from \citet{calette2021_halo} for comparison.}
    \label{fig:mbarm200}
\end{figure}

\begin{figure}
    \centering
    \includegraphics[scale=0.45]{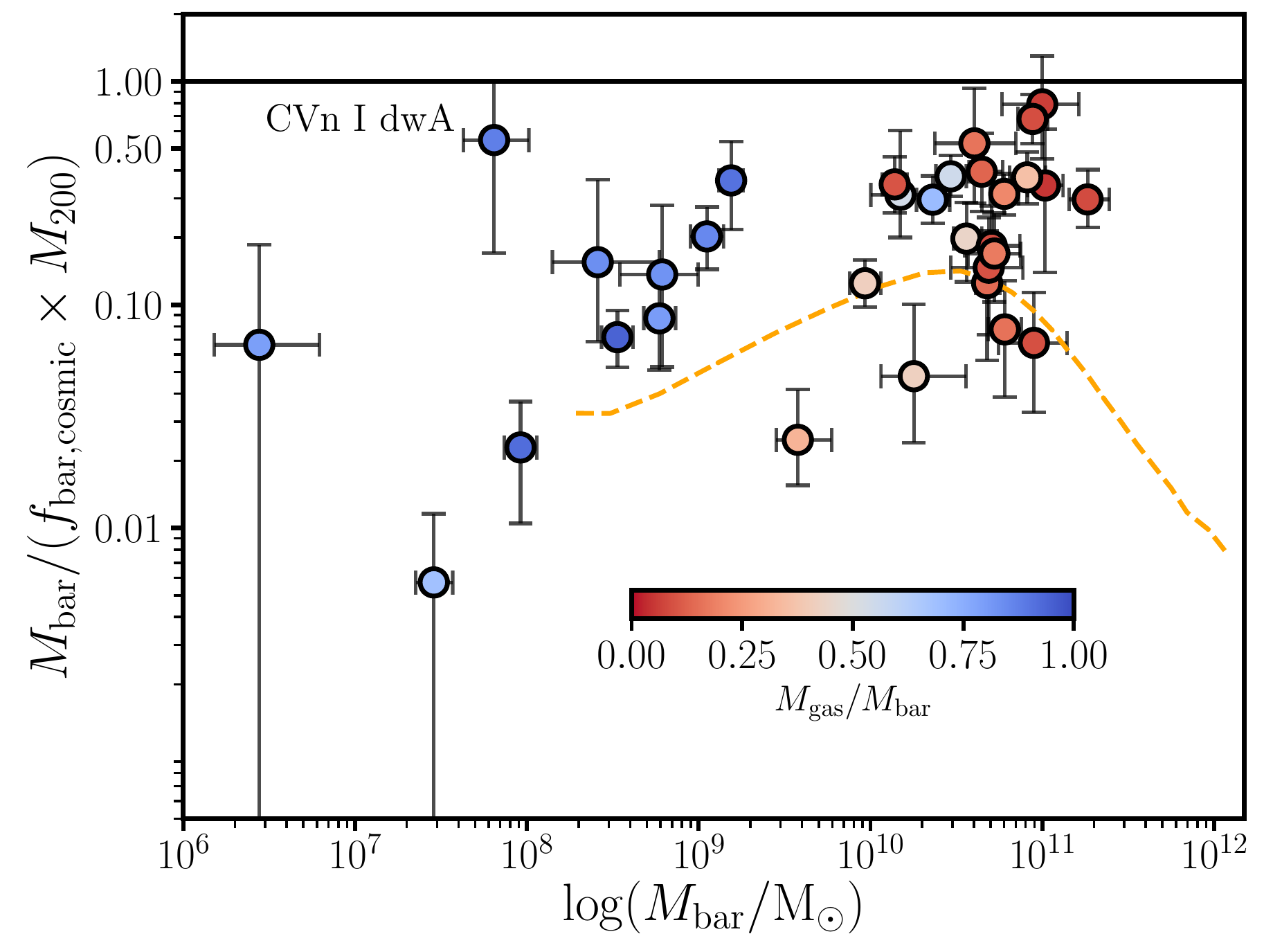}
    \caption{$M_{\rm bar}/M_{200}$ ratio normalised to the cosmological baryon fraction vs. $M_{\rm bar}$. Colours and curves are as in Fig.~\ref{fig:mbarm200}.}
    \label{fig:fbarmbar}
\end{figure}

The top panel of Fig.~\ref{fig:mbarm200} shows the $M_{\rm bar}-M_{200}$ relation for our galaxy sample, while the bottom panel shows the $M_{\rm bar}/M_{200}$ ratio normalised to the cosmic mean. Again, we show the results obtained with the prior on Eq.~\ref{eq:c200M200}. As expected, the picture of the BHMR does not change with respect to the SHMR for the massive galaxies, dominated by the stellar component. Instead, there are substantial changes for the dwarf galaxies once their major baryonic budget is added. Overall, $M_{\rm bar}/M_{200}$ has significant scatter and does not show clear trends as a function of $M_{200}$. A similar behavior is seen in Fig.~\ref{fig:fbarmbar} but as a function of $M_{\rm bar}$.

For comparison, in Figs.~\ref{fig:mbarm200} and \ref{fig:fbarmbar} we plot the BHMR from \citet{calette2021_halo}, which is based on abundance matching to link $M_\ast$ with $M_{200}$ and in empirical correlations to link $M_\ast$ to $M_{\rm gas}$; in this case they consider $M_{\rm gas} = 1.4 M_{\rm HI}$. Predictions from similar semi-empirical approaches overlap with each other \citep{aldo2011}. The relation follows approximately the trends described by the data in the $M_{\rm bar}/(f_{\rm bar,cosmic}\times M_{200})$ vs. $M_{200}$ plane, although at all $M_{200}$ it appears to have lower $M_{\rm bar}$ values than most points. Similarly to what happens in the stellar relation, abundance matching does not seem to capture fully the behavior of the $M_{\rm bar}/(f_{\rm bar,cosmic}\times M_{200})$ vs. $M_{\rm bar}$ plane.\\

\noindent
Finally, we note the peculiar position of the galaxy CVn~I~dwA in the above figures. At $M_{200} \approx 10^{8.8}$, CVn~I~dwA has a higher-than-average $M_{\rm bar}/M_{200}$ with a median value of about 50 per cent the cosmological baryon fraction, although the uncertainties are large. CVn~I~dwA, is a low surface brightness dwarf with an extremely low metallicity \citep{cvnidwa_metals}. It would be interesting to further study this galaxy in the future, in the context of recent results that suggest the existence of a number of extended diffuse dwarf galaxies with baryon fractions comparable to the cosmological average (e.g. \citealt{huds2019,agc114905}).\\

\noindent
Overall, our analysis presents empirical $M_{\rm \ast}-M_{200}$ and $M_{\rm bar}-M_{200}$ relations spanning five orders of magnitude in $M_{\rm \ast}$, $M_{\rm bar}$ and $M_{200}$, taking into account the mass from the atomic and molecular gas, as well as the stellar disc and bulge. These relations can be used to directly test the outcome and predictions from models and simulations, setting constraints on e.g. the gas-to-stellar mass ratio, the efficiency of star formation, and the ability of galaxies to retain their baryons within their discs.

\subsection{Concentration--mass relation}
\label{sec:c200m200}
N-body simulations find a clear anti-correlation between $M_{200}$ and $c_{200}$ (e.g. \citealt{bullock2001,duttonmaccio2014,ludlow2014}). Observational work on rotation curve decomposition, however, reports that this relation is not present in the data unless it is assumed as a prior during the fitting, especially when considering NFW haloes over other more flexible density profiles (see discussion in \citealt{dicintio2014,katz2017,li_massmodels} and references therein). To explore this with the best possible data, in this work we performed our rotation curve decomposition by allowing $c_{200}$ to have 1) an uninformative flat prior, and 2) a prior following the $c_{200}-M_{200}$ of  \citet[][see our Eq.~\ref{eq:c200M200}]{duttonmaccio2014}.

\begin{figure*}
    \centering
    \includegraphics[scale=0.613]{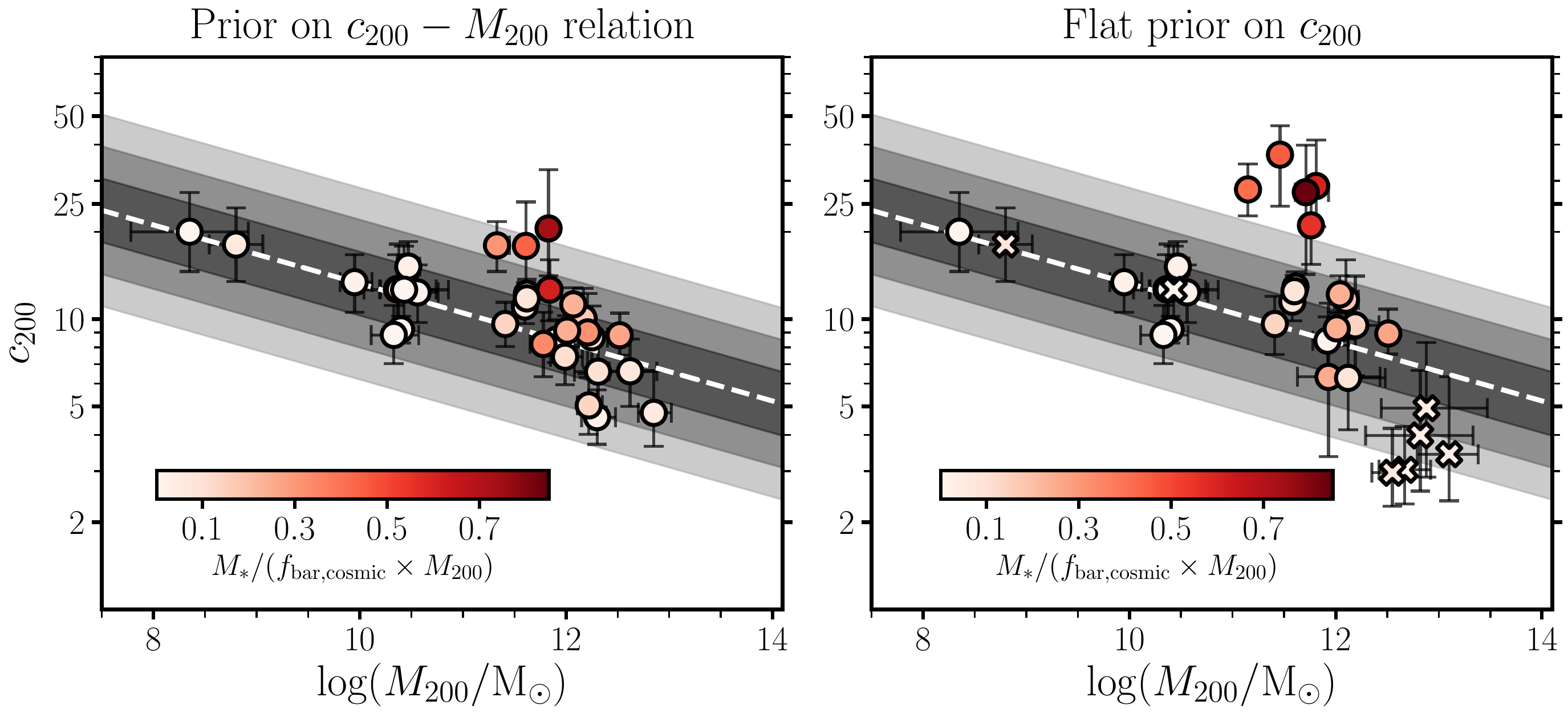}
    \caption{Concentration--mass relation for the dark matter haloes of our galaxy sample. On the \textit{left} panel we display the results considering the prior on Eq.~\ref{eq:c200M200}, while the \textit{right} panel shows the results from the flat prior on $c_{200}$. In both panels we show the $c_{200}-M_{200}$ relation (white line) from the cosmological N-body simulations by \citet{duttonmaccio2014}, together with its 1, 2, 3$\sigma$ scatter (grey bands). Galaxies are colour-coded according to their $M_\ast/(f_{\rm bar,cosmic}\ M_{200}$) fraction. The crosses in the right panel correspond to galaxies for which their mass models are less reliable given their posterior distributions.}
    \label{fig:c200M200_fbar}
\end{figure*}

In Fig.~\ref{fig:c200M200_fbar} we show the location of our galaxies in the $c_{200}-M_{200}$ plane, for both cases of priors on $c_{200}$. As mentioned before, some of our galaxies do not have a well constrained mass model when using a flat prior on $c_{200}$, and so we show them with crosses rather than circles; the values for these galaxies are merely indicative. We compare our measurements against the theoretical expectation from \cite{duttonmaccio2014}, i.e. Eq.~\ref{eq:c200M200}. 

Unsurprisingly, when Eq.~\ref{eq:c200M200} is imposed as a prior (left panel of Fig~~\ref{fig:c200M200_fbar}) we recover very well the $c_{200}-M_{200}$ relation. Remarkably, even when we consider an uninformative prior on $c_{200}$ (right panel of Fig~~\ref{fig:c200M200_fbar}), galaxies appear to follow the $c_{200}-M_{200}$ relation albeit with larger scatter and larger individual uncertainties with respect to the case when Eq.~\ref{eq:c200M200} is used as a prior. Note that the galaxies for which their dark matter halo parameters are less well constrained still follow the same trends when imposing the $c_{200}-M_{200}$ relation on the left panel. Specifically, the most massive galaxies tend to deviate towards low values of $c_{200}$ regardless of the prior.

Our results suggest that the $c_{200}-M_{200}$ relation can actually emerge from the data, provided they are of good enough quality. This is not necessarily expected given that the $c_{200}-M_{200}$ relation is obtained from cosmological N-body simulations, while baryonic physics can modify the structure of dark matter haloes. In particular, adiabatic contraction due to the baryon cooling and collapse, together with stellar/AGN feedback are expected to modify the inner density profiles (and thus the concentration) of the primordial dark matter haloes. In principle, dynamical friction between baryons and dark matter can also affect $c_{200}$, but its effects are expected to be minor for galaxies without extreme bars and where the baryons have retained a large fraction of their angular momentum (\citealt{sellwood2008_review}, see also \citealt{nipoti2015}). Since our sample was chosen not to have significant bars, and the discs have retained a large fraction of the angular momentum of the dark matter halo \citep{paperIBFR}, we expect dynamical friction to be inefficient at affecting $c_{200}$ significantly. Adiabatic contraction tends to increase $c_{200}$, since the halo contracts as response of the deeper potential owed to the central baryon condensation (e.g. \citealt{blumenthal1986,duffy2010,katz2014}). The modification to $c_{200}$ is not straightforward to estimate as it might depend on the individual assembly histories (e.g. \citealt{gnedin2004,tissera2010}). Feedback, on the other hand, mitigates the effects of halo contraction as it can lead to lower central dark matter densities \citep{navarro_cores,read2005,pontzen2012}. This is particularly true if metal-line radiative cooling is efficient (as it eases the collapse of the baryons), but the exact quantification of the effect is highly dependent on the feedback prescriptions (e.g. \citealt{sellwood2008_review,duffy2010,bryan2013,dicintio2014,katz2017}).


Our results indicate that despite the above phenomena, the overall $c_{200}-M_{200}$ relation appears to hold reasonably well for most galaxies in our sample across a mass large range. Recently, \cite{beltz2021} analysed hydrodynamical simulations and obtained similar results. However, simulations also show that the inclusion of baryons can generate `wiggles' in the $c_{200}-M_{200}$ relation due to the interplay of different feedback processes \citep{lovell2018,anbajagane2022}. Interestingly, such deviations seem to be present in our data, at similar $M_{200}$ as in the simulations; namely, the higher concentrations at about $M_{200} \sim 5\times10^{11}~\rm{ M_\odot}$ (where adiabatic contraction is efficient the most), and lower concentrations at $M_{200} \sim 10^{13}~\rm{ M_\odot}$. Note that when we impose the prior of Eq.~\ref{eq:c200M200} on $c_{200}$ the deviations are reduced but still present.

While the comparisons with the results of hydrodynamical simulations are complex and out of the scope of this paper, we can speculate on a qualitative explanation for the trends in our data. With this in mind, in Fig.~\ref{fig:c200M200_fbar} we colour-code the galaxies in terms of their $M_\ast/M_{200}$ fraction normalised to $f_{\rm bar, cosmic}$. As it turns out, the galaxies showing a systematically higher $c_{200}$ are those with the largest $M_\ast/M_{200}$. Based on this, we argue that the galaxies with high $M_\ast/M_{200}$ have larger $c_{200}$ precisely because they converted most of their gas into stars; in fact these galaxies have $M_{\rm bar}/M_{200}$ fractions comparable to the cosmological average host overall higher baryon fractions than galaxies of similar $M_{200}$ (see Fig.~\ref{fig:mbarm200}). 

\citet{postinomissing} has discussed the elevated baryon fractions in some spiral galaxies in detail, arguing for a `failed feedback' scenario where feedback was not efficient enough to eject a significant amount of gas before the galaxies built their stellar mass. In the framework of adiabatic contraction, the high central stellar mass densities could then be responsible for the higher concentrations. Under this scenario, below $M_{200} \sim 5\times10^{11}~\rm{ M_\odot}$ the effects of adiabatic contraction and stellar feedback fully compensate each other (or are both negligible). According to our figure, at $M_{200} \gtrsim 5\times10^{12}~\rm{ M_\odot}$ it would be required that feedback overpowers halo contraction, lowering the concentration and also the global baryon- and stellar-to halo mass fractions. In fact, AGN feedback has been found to be able to lower $c_{200}$ as well as the baryon fraction in galaxies according to simulations (e.g. \citealt{duffy2010,bryan2013}). Moreover, semi-empirical models also find that AGN feedback becomes more efficient at such masses and it manages to reproduce the SHMR (\citealt{marasco2021}, see also e.g. \citealt{fabian2012,choi2015}). Despite the scenario we sketch being merely qualitative, it seems in line with expectations of feedback processes and adiabatic contraction models. However, all this should be tested by repeating the experiment of using a flat prior on $c_{200}$ for larger samples, and ideally complementing it with stellar rotation curves to trace at high resolution the innermost regions of the most massive galaxies, which largely determine their $c_{200}$ parameters. In general, the above trends between $M_{200}$, $c_{200}$, and $M_\ast/M_{200}$ can be exploited to further understand the connection between dark matter haloes and feedback and the build-up of mass in galaxies, providing key constraints to both analytic models and simulations.

\subsection{Comparison with previous work and caveats}
\label{sec:caveats}

\subsubsection{Comparison of mass models with previous works}
Some galaxies in our sample have been recently studied by other authors performing standard rotation curve decomposition. Specifically, \cite{postinomissing} has mass models for seven of our massive galaxies and \cite{li_massmodels} for eight, while \cite{read2017} studied our ten dwarfs. We find reasonably good agreement with all of their results. Below we quantify the comparison. When quoting differences, we refer to the operation of subtracting their values from ours.

First, we compare against \cite{postinomissing}, who fitted NFW haloes using a Gaussian prior on the same $c_{200}-M_{200}$ relation as us, and a flat prior on $\Upsilon_{\rm d}$, and $\Upsilon_{\rm b} = 1.4\Upsilon_{\rm d}$. Given this, for the comparison we use our results from the $c_{200}-M_{200}$ prior. Our mean (median) difference in $M_{200}$ with respect to \cite{postinomissing} is 0.009 ($-0.01$) dex, with a standard deviation of 0.17~dex. Our mean (median) difference in $c_{200}$ is of 0.06 (0.03) dex, with a standard deviation of 0.22~dex (mostly driven by NGC~3992 for which we find a higher $c_{200}$ by 0.5~dex). For $\Upsilon_{\rm d}$ the mean (median) difference is $-0.07$ ($-0.04$) with a standard deviation of 0.08~dex. 

Second, also for the massive galaxies, we compare our fiducial results (i.e. with the flat prior on $c_{200}$) with the values obtained by \cite{li_massmodels}, derived using the same prior on $\Upsilon_{\rm d}$ as us, a flat prior on $c_{200}$, and a prior on $\Upsilon_{\rm b}$ centered at 0.7 with standard deviation of 0.1~dex (which is similar to our prior on $f$). Regarding $M_{200}$, the mean (median) difference is $-0.14$ (0.06) dex with a standard deviation of 0.56~dex, largely driven by NGC~5055 for which our $M_{200}$ is 1.5~dex lower than the Li. et al. value ($0.86\times10^{12}~\rm{ M_\odot}$ vs. $18.6\times10^{12}~\rm{ M_\odot}$). The mean (median) difference in $c_{200}$ is 0.06 (0.09) dex, with a standard deviation of 0.32. The mean (median) difference in $\Upsilon_{\rm d}$ is $-0.03$ ($-0.08$) dex, with standard deviation 0.09~dex. Finally, for $\Upsilon_{\rm b}$ the mean (median) difference is 0.31 (0.32) dex, with standard deviation 0.34~dex. We note here that the bulge-disc decomposition for our sample is more robust than the decomposition used in \cite{li_massmodels}, which is based on a visual fit \citep{sparc}.

Lastly, we compared the results for the dwarfs against the values from \cite{read2017}, who used a flat prior on $c_{200}$ and fitted $M_\ast$ with a Gaussian prior with center and standard deviation according to the values by \cite{zhang}, who obtained $M_\ast$ using SED fitting. The mean (median) difference in $M_{200}$ is 0.1 (0.2) dex, with a standard deviation of 0.34~dex. There seems to be a systematic trend as a function of $M_{200}$, with our more massive galaxies having larger values with respect to \cite{read2017}, but the scatter and low number statistics do not allow to make strong conclusions. For $c_{200}$, the mean (median) difference is 0.1 (0.1) dex, with standard deviation 0.16 dex and our values being lower on average, without clear systematic trends. Overall, taking into account that the approaches of \cite{postinomissing}, \cite{li_massmodels}, and \cite{read2017} differ significantly from ours (e.g. different rotation curves, inclinations, priors, fitting parameters, assumed geometry and shape of the stellar and gas profiles), the agreement in the recovered parameters is fairly good.

\subsubsection{Caveats}

Our mass models and the interpretation of our results are based on a number of assumptions. Here we focus on the two main assumptions that could affect our results the strongest: the premise of vertical hydrostatic equilibrium and the kinematic modelling on which our results rely.

Regarding the former, it is expected that regularly rotating galaxies (as those selected to be in our final sample) without recent strong interactions have their gas pressure in balance with the gravitational pull from the midplane (e.g. \citealt{mo_book,bookFilippo}). Measurements in the Milky Way seem to indicate that some fraction of the atomic gas might have ~50\% larger scale height than expected from hydrostatic equilibrium within $\sim$ 5~kpc, although the agreement improves at larger radii (see \citealt{marasco2017,ceci_MW} for a detailed discussion). In general, in order to test the validity of hydrostatic equilibrium it is of primary interest to measure both the gas velocity dispersion and the gas scale height in other galaxy discs directly from observations, but there is no robust method to do so yet. Note that given our finding that the scale height influences very little the mass models except for the lowest mass galaxies, small deviations from vertical hydrostatic equilibrium would have no impact on our results regarding the SHMR, BHMR, or the $c_{200}-M_{200}$ relation.


Regarding our second assumption, one of the main motivations and strength of the paper is the argument that we self-consistently derive our mass models and scale heights taking into account each other given our iterative approach. However, the kinematic modelling from which our rotation curves and velocity dispersion profiles are derived is not consistent with the fact that gas discs are flared. The 3D kinematic modelling technique of \BB{} fits thin tilted-ring models \citep{rogstad1974} to the data cubes. This represents an improvement over fitting tilted-ring models to the velocity fields, but neither approach takes into account the increase of the scale height with radius. This is because for thick/flared discs, line of sights contain emission coming from different rings, and the ring-by-ring approach of the tilted-ring technique cannot account for this. 


Despite this, the effects of assuming a thin disc when a thick disc is appropriate are known. As discussed in \cite{iorio_phd} and \cite{enrico_smc}, there are three main effects due to the mixing of different disc layers in the same line of sight. By considering a thin disc instead of a thick disc: \textit{i)} the recovered surface density profiles are shallower than in reality, \textit{ii)} the velocity dispersion tends to be overestimated, and \textit{iii)} the rotation velocities at large (low) radii can be slightly overestimated (underestimated). These effects are expected to be negligible for massive spiral galaxies, but somewhat more important for dwarfs, as their potential wells are shallower and their thickness are larger both in their overall value and relative to their $R_{\rm max,HI}$ (see Fig.~\ref{fig:all_h}). Nevertheless, as shown by \cite{iorio_phd}, even for three of the least massive galaxies in our sample (DDO~210, NGC~2366, and WLM) the kinematic parameters derived with \BB{} are well consistent with the values estimated considering a thick disc, and the flaring is a subdominant source of uncertainty in the kinematic modelling compared to, for instance, instrumental limitations and non-circular motions. All of the above indicates that the kinematic parameters derived with the tilted-ring modelling are reliable, even if this technique cannot fully take into account the flaring of discs. In conclusion, this partial inconsistency in our methodology does not significantly affect our results.


\section{Conclusions}
\label{sec:conclusions}
Mass models of galaxies obtained through their rotation curve decomposition are often derived assuming that the neutral (HI) and molecular (H$_2$) gas discs are razor-thin. In reality, gas discs are flared, which can affect the mass models given the different gravitational effects of a flared geometry with respect to a razor-thin disc. In this paper we set out to investigate the effects of the flaring on the best-fitting parameters from rotation curve decomposition, namely the stellar mass-to-light ratios ($\Upsilon$), the mass of the halo ($M_{200}$), and its concentration ($c_{200}$). This is the first time that such an investigation is carried out systematically.

To this end, we use a sample of 32 late-type galaxies with the best possible data quality. Our sample has H\,{\sc i} and CO resolved interferometric observations, bulge-disc decomposition based on deep NIR imaging, and available 3D kinematic modelling providing their gas velocity dispersion and circular speed ($V_{\rm c}$). The sample spans about six orders of magnitude in stellar mass ($M_\ast$), including very small ($V_{\rm c} \approx 20\ \rm{km/s}$) and very massive ($V_{\rm c} \approx 290\ \rm{km/s}$) galaxies. Under the assumption of vertical hydrostatic equilibrium we use an iterative approach combined with a Bayesian fitting technique to simultaneously derive the scale height of the gas discs and the mass models (using NFW and \textsc{core}NFW profiles). The NFW and \textsc{core}NFW profiles provide an excellent fit to the observations of our massive and dwarf galaxies, respectively. All our mass models converge with reasonable values of $\Upsilon$, $M_{200}$ and $c_{200}$ (Fig.~\ref{fig:mmgau}). For most of our galaxies the fits are equally satisfactory by fitting $c_{200}$ using a flat prior or by imposing a reference $c_{200}-M_{200}$ relation from N-body cosmological simulations, although for some of them the $c_{200}-M_{200}$ relation is necessary to set robust constraints on the mass models.
In addition to the derivation of the gas discs flaring and the analysis of its effects, we use our detailed mass models to explore the stellar--to--halo (SHMR), baryon--to--halo (BHMR), and $c_{200}-M_{200}$ relations. Our main results can be summarised as follows.

\begin{itemize}
    \item In the majority of our sample the effects of the flaring of the gas discs on the recovered best-fitting parameters is very minor given the larger dynamical contribution of the stellar and dark matter components. However, the effects are significant for the smallest and gas-richest dwarf galaxies, where the lower gas circular speeds of a flared disc gives room to considerably more massive haloes (Fig.~\ref{fig:mmgau}, \ref{fig:cvidwa}, \ref{fig:comparison}). 
    \item The scale heights of both H\,{\sc i} and CO discs are monotonically increasing with radius in the vast majority of cases. As expected from their lower gas velocity dispersion, CO discs are considerably thinner than the H\,{\sc i} discs. Additionally, low-mass galaxies have thicker discs than more massive ones (Fig.~\ref{fig:example_h} and \ref{fig:all_h}). We find little evidence of a `universal' scale height profile (Fig.~\ref{fig:all_h}), contrary to previous claims.
    \item We explore the stellar--to--halo mass relation for our sample (Fig.~\ref{fig:mstarm200}). At fixed $M_{200}$, galaxies with higher stellar-to-gas ratios ($M_\ast/M_{\rm gas}$) have higher $M_\ast/M_{200}$. Some galaxies around $M_{200} \approx 10^{12}~\rm{ M_\odot}$ have very high $M_\ast/M_{200}$ ratios approaching the cosmological baryon fraction, in agreement with previous findings. Nearly at all masses the galaxies scatter between a linear monotonically increasing relation and a broken power-law following abundance-matching expectations. As a function of $M_\ast$, the $M_\ast/M_{200}$ ratios increase nearly monotonically (Fig.~\ref{fig:fstarmstar}).
    \item The baryon--to--halo mass relation of our sample (Fig.~\ref{fig:mbarm200}) is similar to the stellar--to--halo mass relation at high baryonic masses ($M_{\rm bar}$) but it is shallower in the low-mass regime given the high gas masses in dwarf galaxies. In general, the $M_{\rm bar}/M_{200}$ ratio of is more homogeneous, though with large scatter, than $M_\ast/M_{200}$ (Figs.~\ref{fig:mbarm200} and \ref{fig:fbarmbar}).
    \item The $c_{200}-M_{200}$ relation of our dark matter haloes follows well the expected relation from N-body cosmological simulations. This is trivial when the $c_{200}-M_{200}$ relation is imposed, but the relation appears to hold relatively well even when using an uninformative prior on $c_{200}$ (Fig.~\ref{fig:c200M200_fbar}). That galaxies follow the $c_{200}-M_{200}$ relation is not necessarily expected given that adiabatic contraction and feedback processes can modify the inner distribution of the dark matter density profile.
    \item Despite the overall $c_{200}-M_{200}$ relation being preserved, galaxies with $M_{200} \approx 5\times 10^{11}~\rm{ M_\odot}$ and $M_{200} \approx 10^{13}~\rm{ M_\odot}$ seem to lie systematically above and below the relation, respectively. Interestingly, the galaxies above the relation show also the highest stellar and baryon mass fractions, opposite to the galaxies below the relation (Fig.~\ref{fig:c200M200_fbar}). We argue that these deviations could be attributed to adiabatic contraction and AGN feedback, respectively.
\end{itemize}

\noindent
We provide the largest sample for which the gas scale heights have been derived homogeneously using the hydrostatic equilibrium condition, allowing us to derive most detailed mass models. Our analysis is relevant in the context of the H\,{\sc i} kinematics and mass modeling from the upcoming large-volume H\,{\sc i} surveys. Moreover, our results can be directly used to test the outcome and predictions of both models and hydrodynamical simulations, with the goal of achieving a more complete picture of the physics regulating the shape of galaxies as well as the galaxy-halo connection and the efficiency of feedback processes.

\section*{Acknowledgements}

We thank Heikki Salo for different clarifications on the S4G bulge-disc decomposition, and Eduardo Vitral for useful discussions regarding the reprojection of surface brightness profiles. We also thank Toby Brown for sharing with us the VERTICO data for NGC~4651 and NGC~4698, Hong-Xin Zhang for the surface brightness profile of LITTLE THINGS galaxies, Brad Frank for the comparison CO profiles of a few galaxies in our sample, and Karin Sandstrom for the $\alpha_{_{\rm CO}}$ profiles of some of our galaxies. The comments from an anonymous referee are also appreciated. P.E.M.P. and F.F. are supported by the Netherlands Research School for Astronomy (Nederlandse Onderzoekschool voor Astronomie, NOVA), research programme Network 1, Project 10.1.5.6. E.A.K.A. is supported by the WISE research programme, which is financed by the Dutch Research Council (NWO).
We have used extensively SIMBAD \citep{simbad}, NED, and ADS services, as well the Python packages NumPy \citep{numpy}, Matplotlib \citep{matplotlib}, SciPy \citep{scipy}, and Astropy \citep{astropy}, and the tool TOPCAT \citep{topcat} for which we are thankful.

\section*{Data Availability}

All the data used to derive our mass models are public or available in the references and links provided in the main text. Our mass models (rotation curve decompositions and MCMC posterior distributions) and scale heights can be found \href{https://www.dropbox.com/sh/768phag6ttgym0j/AABZwerBgFofkkZ5npHW17YBa?dl=0}{here}. Any other data product derived from this paper is available upon request to the authors.



\bibliographystyle{mnras}
\bibliography{references} 



\begin{appendix}
\section{Mass models with a flat prior on $c_{200}$}
Table~\ref{tab:results_flat} lists the best-fitting parameters of our rotation curve decomposition obtained using a flat prior on $c_{200}$.



\begin{table*}
\caption{Results of our rotation curve decomposition for the case when $c_{200}$ has a flat prior. The columns give our fiducial values (the median of the posterior distributions) and their upper and lower uncertainties (corresponding to the $16^{\rm th}$ and $84^{\rm th}$ percentiles, including the contribution from distance uncertainties. Only galaxies with bulges have values of $f$. Note that the posterior distributions of the galaxies CVn I dwA, DDO~210, NGC~253, NGC~3486, NGC~3898, NGC~4535, NGC~5350, and UGC~8505, are not Gaussian, and thus the reported percentiles are not as reliable as in the case of the prior on Eq.~\ref{eq:c200M200}, but we report them here for completeness.}
\label{tab:results_flat}
\begin{center} 
\begin{tabular}{lcccccccccccc}
	\hline \noalign{\smallskip}
	\multicolumn{1}{c}{Name} & \multicolumn{3}{c}{$\log(M_{200}/\rm{M_\odot})$} & \multicolumn{3}{c}{$c_{200}$} & \multicolumn{3}{c}{$\Upsilon_{\rm d} (\rm{M_\odot/L_\odot})$} & \multicolumn{3}{c}{$f$} \\ \noalign{\smallskip} 
 & $50^{\rm th}$ pctl & $\sigma-$ & $\sigma+$ & $50^{\rm th}$ pctl & $\sigma-$ & $\sigma+$ & $50^{\rm th}$ pctl & $\sigma-$ & $\sigma+$ & $50^{\rm th}$ pctl & $\sigma-$ & $\sigma+$  \\ \noalign{\smallskip} \hline
CVn I dwA & 8.82 & $-$0.40 & + 0.74 & 17.19 & $-$11.74 & + 19.64 & 0.50 & $-$0.11 & + 0.11 & -- & -- & -- \\ 
DDO 52 & 10.30 & $-$0.26 & + 0.51 & 14.07 & $-$7.25 & + 9.51 & 0.50 & $-$0.11 & + 0.11 & -- & -- & -- \\ 
DDO 87 & 10.06 & $-$0.17 & + 0.46 & 25.10 & $-$15.93 & + 13.32 & 0.51 & $-$0.11 & + 0.11 & -- & -- & -- \\ 
DDO 126 & 9.99 & $-$0.21 & + 0.37 & 12.56 & $-$4.81 & + 5.30 & 0.46 & $-$0.11 & + 0.11 & -- & -- & -- \\ 
DDO 154 & 10.44 & $-$0.09 & + 0.10 & 8.82 & $-$0.99 & + 1.07 & 0.45 & $-$0.11 & + 0.11 & -- & -- & -- \\ 
DDO 168 & 10.52 & $-$0.20 & + 0.30 & 13.06 & $-$4.06 & + 4.41 & 0.43 & $-$0.11 & + 0.11 & -- & -- & -- \\ 
DDO 210 & 8.36 & $-$0.64 & + 0.92 & 19.05 & $-$11.18 & + 16.86 & 0.50 & $-$0.11 & + 0.11 & -- & -- & -- \\ 
NGC 0253 & 12.82 & $-$0.53 & + 0.51 & 3.98 & $-$1.42 & + 2.69 & 0.34 & $-$0.03 & + 0.03 & 1.40 & $-$0.10 & + 0.10 \\ 
NGC 1313 & 11.92 & $-$0.14 & + 0.20 & 8.44 & $-$1.75 & + 1.75 & 0.31 & $-$0.09 & + 0.09 & 1.38 & $-$0.10 & + 0.10 \\ 
NGC 2366 & 10.38 & $-$0.09 & + 0.12 & 18.81 & $-$4.38 & + 4.52 & 0.49 & $-$0.11 & + 0.11 & -- & -- & -- \\ 
NGC 2403 & 11.58 & $-$0.06 & + 0.07 & 11.49 & $-$1.62 & + 1.81 & 0.40 & $-$0.07 & + 0.07 & -- & -- & -- \\ 
NGC 2841 & 12.51 & $-$0.06 & + 0.05 & 8.93 & $-$1.35 & + 1.90 & 0.83 & $-$0.07 & + 0.08 & 1.38 & $-$0.09 & + 0.09 \\ 
NGC 3198 & 11.61 & $-$0.02 & + 0.03 & 12.93 & $-$1.53 & + 1.66 & 0.37 & $-$0.07 & + 0.07 & -- & -- & -- \\ 
NGC 3351 & 11.93 & $-$0.30 & + 0.50 & 6.33 & $-$2.96 & + 5.02 & 0.59 & $-$0.06 & + 0.05 & 1.40 & $-$0.09 & + 0.09 \\ 
NGC 3486 & 12.67 & $-$0.25 & + 0.25 & 3.03 & $-$0.72 & + 1.25 & 0.39 & $-$0.08 & + 0.10 & 1.40 & $-$0.10 & + 0.10 \\ 
NGC 3621 & 11.60 & $-$0.05 & + 0.05 & 12.49 & $-$1.59 & + 1.69 & 0.20 & $-$0.05 & + 0.05 & -- & -- & -- \\ 
NGC 3675 & 12.19 & $-$0.17 & + 0.23 & 9.53 & $-$3.14 & + 4.55 & 0.43 & $-$0.08 & + 0.11 & -- & -- & -- \\ 
NGC 3898 & 13.10 & $-$0.31 & + 0.28 & 3.43 & $-$1.06 & + 2.91 & 0.67 & $-$0.11 & + 0.16 & 1.42 & $-$0.11 & + 0.11 \\ 
NGC 3992 & 11.81 & $-$0.05 & + 0.06 & 28.77 & $-$7.96 & + 12.64 & 0.60 & $-$0.10 & + 0.10 & 1.39 & $-$0.10 & + 0.10 \\ 
NGC 4535 & 12.88 & $-$0.44 & + 0.59 & 4.94 & $-$2.08 & + 3.38 & 0.56 & $-$0.10 & + 0.08 & 1.44 & $-$0.10 & + 0.10 \\ 
NGC 4536 & 12.12 & $-$0.23 & + 0.36 & 6.28 & $-$2.12 & + 2.34 & 0.29 & $-$0.04 & + 0.05 & 1.36 & $-$0.10 & + 0.10 \\ 
NGC 4559 & 11.41 & $-$0.07 & + 0.09 & 9.63 & $-$2.06 & + 2.37 & 0.36 & $-$0.09 & + 0.09 & 1.40 & $-$0.10 & + 0.10 \\ 
NGC 4651 & 11.46 & $-$0.07 & + 0.08 & 36.89 & $-$12.36 & + 9.46 & 0.45 & $-$0.13 & + 0.12 & 1.42 & $-$0.10 & + 0.10 \\ 
NGC 4698 & 12.10 & $-$0.11 & + 0.12 & 11.68 & $-$3.14 & + 4.32 & 0.45 & $-$0.08 & + 0.09 & 1.38 & $-$0.10 & + 0.10 \\ 
NGC 4725 & 11.76 & $-$0.05 & + 0.06 & 21.06 & $-$5.63 & + 6.07 & 0.51 & $-$0.10 & + 0.10 & 1.38 & $-$0.10 & + 0.10 \\ 
NGC 4736 & 11.15 & $-$0.07 & + 0.08 & 28.00 & $-$5.30 & + 6.28 & 0.18 & $-$0.03 & + 0.03 & 1.35 & $-$0.10 & + 0.10 \\ 
NGC 5005 & 11.71 & $-$0.10 & + 0.22 & 27.39 & $-$13.12 & + 12.41 & 0.41 & $-$0.08 & + 0.08 & 1.38 & $-$0.10 & + 0.10 \\ 
NGC 5033 & 12.04 & $-$0.05 & + 0.05 & 12.18 & $-$1.70 & + 1.92 & 0.30 & $-$0.05 & + 0.07 & 1.41 & $-$0.10 & + 0.10 \\ 
NGC 5055 & 12.01 & $-$0.04 & + 0.04 & 9.30 & $-$1.17 & + 1.31 & 0.37 & $-$0.04 & + 0.04 & -- & -- & -- \\ 
NGC 5350 & 12.55 & $-$0.20 & + 0.19 & 2.96 & $-$0.69 & + 1.25 & 0.53 & $-$0.08 & + 0.11 & 1.46 & $-$0.10 & + 0.10 \\ 
UGC 8508 & 9.80 & $-$0.40 & + 0.88 & 23.54 & $-$13.45 & + 15.03 & 0.50 & $-$0.11 & + 0.11 & -- & -- & -- \\ 
WLM & 10.70 & $-$0.34 & + 0.38 & 6.27 & $-$1.70 & + 2.25 & 0.49 & $-$0.11 & + 0.11 & -- & -- & -- \\ 
   \hline
    \end{tabular}
  \end{center}
\end{table*}

\section{Baryon content of our galaxy sample}
Table~\ref{tab:masses} presents the estimates of the visible mass in our galaxies. The mass of the gas includes H\,{\sc i} and H$_2$ (when available), while the stellar mass includes the contribution of disc and bulge (when available). The baryonic mass is the sum of the gas and stellar mass. Note that $M_\ast$ has been derived using the mass-to-light ratios obtained with the prior on Eq.~\ref{eq:c200M200}.

\begin{table*}
\caption{Stellar, gas, and baryonic (stars+gas) mass of our galaxy sample. The stellar mass has been derived using the mass-to-light ratios from our mass models considering the prior on the $c_{200}-M_{200}$ relation.}
\label{tab:masses}
\begin{center} 
\begin{tabular}{lccccccccc}
	\hline \noalign{\smallskip}
	\multicolumn{1}{c}{Name} &   \multicolumn{3}{c}{$\log(M_{\ast}/\rm{M_\odot})$} & \multicolumn{3}{c}{$\log(M_{\rm gas}/\rm{M_\odot})$} & \multicolumn{3}{c}{$\log(M_{\rm bar}/\rm{M_\odot})$} \\ \noalign{\smallskip} 
 & $50^{\rm th}$ pctl & $\sigma-$ & $\sigma+$ & $50^{\rm th}$ pctl & $\sigma-$ & $\sigma+$ & $50^{\rm th}$ pctl & $\sigma-$ & $\sigma+$ \\ \noalign{\smallskip} \hline
CVn I dwA &  6.88 & -0.13 & 0.12 &  7.76 & -0.21 & 0.22 &  7.81 &  -0.18 & 0.20 \\
DDO 52    &  8.14 & -0.13 & 0.12 &  9.15 & -0.08 & 0.08 &  9.19 &  -0.07 & 0.07 \\
DDO 87    &  8.02 & -0.32 & 0.26 &  8.71 & -0.31 & 0.24 &  8.79 &  -0.25 & 0.21 \\
DDO 126   &  7.61 & -0.15 & 0.13 &  8.34 & -0.34 & 0.40 &  8.41 &  -0.26 & 0.36 \\
DDO 154   &  7.24 & -0.14 & 0.12 &  8.50 & -0.10 & 0.10 &  8.53 &  -0.09 & 0.09 \\
DDO 168   &  8.06 & -0.15 & 0.13 &  8.68 & -0.11 & 0.11 &  8.77 &  -0.09 & 0.09 \\
DDO 210   &  5.75 & -0.14 & 0.12 &  6.34 & -0.36 & 0.41 &  6.44 &  -0.26 & 0.35 \\
NGC 0253  & 10.61 & -0.08 & 0.08 &  9.85 & -0.15 & 0.20 & 10.68 &  -0.07 & 0.08 \\
NGC 1313  &  9.41 & -0.13 & 0.13 &  9.09 & -0.35 & 0.42 &  9.58 &  -0.13 & 0.20 \\
NGC 2366  &  8.20 & -0.15 & 0.13 &  8.98 & -0.11 & 0.11 &  9.05 &  -0.10 & 0.10 \\
NGC 2403  &  9.73 & -0.13 & 0.12 &  9.59 & -0.13 & 0.14 &  9.97 &  -0.09 & 0.09 \\
NGC 2841  & 11.22 & -0.12 & 0.11 & 10.21 & -0.41 & 0.52 & 11.26 &  -0.11 & 0.13 \\
NGC 3198  & 10.12 & -0.13 & 0.12 & 10.21 & -0.13 & 0.13 & 10.47 &  -0.09 & 0.09 \\
NGC 3351  & 10.59 & -0.08 & 0.08 &  9.77 & -0.30 & 0.48 & 10.65 &  -0.07 & 0.12 \\
NGC 3486  & 10.01 & -0.23 & 0.2  &  9.88 & -0.43 & 0.51 & 10.25 &  -0.19 & 0.30 \\
NGC 3621  &  9.83 & -0.14 & 0.12 & 10.21 & -0.13 & 0.12 & 10.36 &  -0.09 & 0.10 \\
NGC 3675  & 10.64 & -0.25 & 0.21 &  9.69 & -0.24 & 0.37 & 10.69 &  -0.22 & 0.20 \\
NGC 3898  & 10.91 & -0.23 & 0.21 &  9.93 & -0.29 & 0.23 & 10.95 &  -0.21 & 0.19 \\
NGC 3992  & 10.97 & -0.25 & 0.21 &  9.80 & -0.61 & 0.69 & 11.00 &  -0.23 & 0.21 \\
NGC 4535  & 10.70 & -0.11 & 0.11 & 10.00 & -0.11 & 0.11 & 10.78 &  -0.10 & 0.10 \\
NGC 4536  & 10.30 & -0.10 & 0.10 & 10.21 & -0.12 & 0.14 & 10.56 &  -0.08 & 0.09 \\
NGC 4559  &  9.84 & -0.14 & 0.13 &  9.91 & -0.35 & 0.43 & 10.17 &  -0.17 & 0.28 \\
NGC 4651  & 10.53 & -0.28 & 0.24 &  9.83 & -0.43 & 0.55 & 10.60 &  -0.23 & 0.24 \\
NGC 4698  & 10.67 & -0.18 & 0.17 &  9.65 & -0.26 & 0.43 & 10.71 &  -0.16 & 0.16 \\
NGC 4725  & 10.90 & -0.09 & 0.09 &  9.90 & -0.16 & 0.16 & 10.94 &  -0.08 & 0.09 \\
NGC 4736  & 10.09 & -0.08 & 0.08 &  9.15 & -0.11 & 0.12 & 10.14 &  -0.07 & 0.07 \\
NGC 5005  & 10.99 & -0.11 & 0.11 &  9.75 & -0.18 & 0.38 & 11.02 &  -0.10 & 0.11 \\
NGC 5033  & 10.72 & -0.14 & 0.14 & 10.47 & -0.15 & 0.15 & 10.91 &  -0.10 & 0.11 \\
NGC 5055  & 10.68 & -0.10 & 0.10 & 10.10 & -0.10 & 0.10 & 10.78 &  -0.08 & 0.08 \\
NGC 5350  & 10.63 & -0.19 & 0.17 &  9.98 & -0.25 & 0.24 & 10.72 &  -0.15 & 0.15 \\
UGC 8508  &  6.95 & -0.15 & 0.14 &  7.30 & -0.14 & 0.14 &  7.46 &  -0.10 & 0.11 \\
WLM       &  6.85 & -0.13 & 0.12 &  7.93 & -0.10 & 0.10 &  7.96 &  -0.09 & 0.09 \\
   \hline
    \end{tabular}
  \end{center}
\end{table*}
\end{appendix}


\bsp	
\label{lastpage}
\end{document}